\documentclass[12pt,a4paper]{article}
\usepackage{amsmath}
\usepackage{amssymb}
\title{Det-Det Correlations for Quantum Maps: Dual Pair and
  Saddle-Point Analyses} 
\author{S. Nonnenmacher\thanks{Service de Physique Th\'eorique, CEA,
91191 Gif-sur-Yvette, France} and M.R. Zirnbauer\thanks{Institut f\"ur Theoretische Physik, Universit\"at zu K\"oln, Z\"ulpicherstr. 77, 50937 K\"oln, Germany}}
\date{}

\begin{document}

\def\ad{\mathop{\rm ad}\nolimits}
\def\Ad{\mathop{\rm Ad}\nolimits}
\def\Det{\mathop{\rm Det}\nolimits}
\def\un{\mathop{\mathfrak{u}(N)}\nolimits}
\def\eps{\epsilon}
\def\vareps{\varepsilon}

\def\Ker{\mathop{\rm Ker}\nolimits}
\def\tr{{\rm tr}}
\def\Tr{{\rm Tr}}
\def\mi{{\rm i}}
\def\e{\mathop{\rm e}\nolimits}
\def\sq2{\sqrt{2}}
\def\defi{\stackrel{\rm def}{=}}
\def\t2{{\mathbb T}^2}
\def\tc{T_{\mathbb C}}
\def\s2{{\mathbb S}^2}
\def\hn{{\cal H}_{N,\kappa}}
\newcommand {\Bl} {\big\langle}
\newcommand {\Br} {\big\rangle}

\maketitle 

{\abstract An attempt is made to clarify the ballistic non--linear
  sigma model formalism recently proposed for quantum chaotic systems,
  by looking at the spectral determinant $Z(s) = {\rm Det}(1-sU)$ for
  quantized maps $U \in {\rm U}(N)$, and studying the correlator
  $\omega_U(s) = \int d\theta \, |Z({\rm e}^{{\rm i}\theta} s)|^2$.
  By identifying ${\rm U}(N)$ as one member of a dual pair acting in
  the spinor representation of ${\rm Spin}(4N)$, the expansion of
  $\omega_U(s)$ in powers of $s^2$ is shown to be a decomposition into
  irreducible characters of ${\rm U}(N)$.  In close analogy with the
  ballistic non--linear sigma model, a coherent--state integral
  representation of $\omega_U (s)$ is developed.  For generic $U$ this
  integral has $\binom{2N}{N}$ saddle points and the leading--order
  saddle--point approximation turns out to reproduce $\omega_U(s)$
  \emph{exactly}, up to a constant factor.  This miracle is explained
  by interpreting $\omega_U(s)$ as a character of ${\rm U}(2N)$, and
  arguing that the leading--order saddle--point result corresponds to
  the \emph{Weyl character formula}.  Unfortunately, the Weyl
  decomposition behaves non--smoothly in the semiclassical limit $N
  \to \infty$, and to make further progress some additional averaging
  needs to be introduced.  Several schemes are investigated, including
  averaging over basis states and an ``isotropic'' average.  The
  saddle--point approximation applied in conjunction with these
  schemes is demonstrated to give incorrect results in general, one
  notable exception being a semiclassical averaging scheme, for which
  all loop corrections vanish identically.  As a side product of the
  dual pair decomposition with isotropic averaging, the crossover
  between the Poisson and CUE limits is obtained.}

\section{Introduction}

One of the striking characteristics of a quantized chaotic Hamiltonian
system is found in the correlations inherent in its spectrum at small
energy differences.  Extensive numerical work has shown that various
quantities (such as the nearest-neighbour spacing distribution and the
two--level correlation function) of a quantum chaotic system are
\emph{universal}: their behaviour coincides with that of a
Wigner-Dyson random matrix ensemble of the appropriate symmetry class
\cite{bohigas}.  This property, first noticed in billiards, was found
to apply to many chaotic systems, including symplectic \emph{maps}.
In contrast, if the dynamics is integrable (in the sense that the
$2f$--dimensional phase space foliates into $f$--dimensional
submanifolds invariant under the Hamiltonian flow) the generic
behaviour of the eigenvalues is expected \cite{berrytabor} to be that
of independent random variables, so that their correlations are in the
Poisson universality class.

The present paper will be concerned with quantum maps, i.e.~with
quantizations of some canonical transformation $\phi : M \to M$ of a
compact symplectic manifold $M$.  We assume that the problem of
quantization itself has been tackled, so the phase space has been
prequantized into a Hilbert space ${\cal H}_N$ of dimension $N \sim
\hbar^{-1}$, and the quantum map acts on it as a unitary operator
\cite{tabor,zeld2}.  With respect to a basis of ${\cal H}_N$ this
operator is represented by an $N\times N$ unitary matrix $U_{\phi,N}$.
The latter has a semiclassical limit, in the sense that traces of its
powers can be estimated in terms of classical periodic points
\cite{tabor}.  For a system with one degree of freedom, the
Gutzwiller--Tabor trace formula reads
\begin{equation}\label{gutzwiller}
  {\rm Tr} (U_{\phi,N}^n)\stackrel{N\to\infty}{\sim}\sum_{p\subset
    {\rm Fix}(\phi^n)} N^{{\rm dim}(p)/2} A_p \e^{\mi N\Phi p} \;,
\end{equation}
where $p$ is one component of the set of $n$--periodic points; for an
Anosov system, it is an isolated point (${\rm dim} (p) = 0$), whereas
if the dynamics conserves energy, $p$ is 1-dimensional.  $\Phi_p$ and
$A_p$ are purely classical quantities related to the dynamics around
the set $p$.

The quantum spectrum consists of the $N$ eigenvalues
(pseudo--energies) $\{\e^{\mi\theta_j}\}_{j=1,\ldots,N}$ of $U_{\phi,
  N}$.  The first analytical estimates of the two--level correlation
function (which is the Fourier transform of the \emph{form factor}
$F(n) = |{\rm Tr}(U_{\phi, N}^n) |^2$) for such spectra were based on
the above trace formula, combined with some known ergodic properties
of long periodic orbits \cite{berry85}.  In the present paper we focus
attention on another statistic, namely the autocorrelation function of
the spectral determinant:
\begin{equation}\label{omegaU}
  \Omega_U(\gamma) \defi \gamma^{-N/2} \int_0^{2\pi} \frac{d\phi}
  {2\pi}\ {\rm Det}(1-\gamma\e^{\mi\phi}U){\rm Det} (1 - \e^{-\mi\phi}
  U^\dagger)\;.
\end{equation}
(The parameter $\gamma$ will be a complex number close to unity, with
the scaling $|\gamma - 1| \sim 1/N$.)  This correlation function has
already been considered \cite{haake,smilansky} for chaotic versus
integrable quantum maps, and the same universality was observed as for
the form factor or the nearest-neighbour distribution.  A
semiclassical analysis of this correlation function was performed
using the Gutzwiller trace formula in \cite{smilansky}.

The computation of correlation functions from the trace formula
(\ref{gutzwiller}) always requires some sort of averaging.  In the
semiclassical theory of the form factor $F(n)$ one wants to use the
so-called diagonal approximation, neglecting the off-diagonal terms in
the double-sum over periodic orbits \cite{berry85}.  To justify this
step one must average over energy or some family of systems: one needs
slight variations in the classical actions $\Phi_p$ to make the phase
interferences $\sim \e^{\mi N( \Phi_{p}-\Phi_{p'})}$ average to zero.
The need for averaging was emphasized in \cite{haake,prange}, where it
was pointed out that the spectrum of an individual quantum system is
too noisy to allow universality to be seen in its bare form factor.
In addition to the noise problem, there exist some quantum chaotic
systems with arithmetic symmetries, which lead to periodic orbit
degeneracies and non--universal spectral correlations
\cite{bogoleyvraz,mezzadri}.  (Such systems are non--generic, however,
in any decent space of smooth maps.)

Thus universal behaviour is expected only in the generic case, and to
make a correct mathematical statement about universality of the
spectral correlations of a general system one ought to define the
precise meaning of the word ``generic''.  (In the case of integrable
systems, the spectral correlations could sometimes be studied
directly, by utilizing the explicit expressions for the eigenvalues;
two--point correlations were shown to be Poisson for a rather subtle
set of parameters \cite{marklof}.)  We can avoid the issue of
genericity by averaging the correlation function over some set of
quantum maps.  That is, we specify a measure ${\rm d}{\cal P}_N(U)$ on
the unitary group ${\rm U}(N)$, and the function to be studied then
reads
\begin{displaymath}
  \Bl \Omega_U (\gamma)\Br = \int_{{\rm U}(N)}{\rm d}{\cal P}_N(U) \;
  \Omega_U(\gamma) \;.
\end{displaymath}
We want this measure to be very concentrated (or ``local'') around the
quantum map $U_{\phi,N}$ (see Sections \ref{semiclas},
\ref{HeatKernel}).  In the course of this article, we will also
consider cases for which this measure has a broader support (Section
\ref{eigenbasisave}), including the case where the measure is the Haar
measure on ${\rm U}(N)$.  We are then dealing with the circular
unitary ensemble (CUE), for which the determinant correlation function
has been thoroughly investigated \cite{haake,zirnbauer96}.

The Gutzwiller trace formula has the attractive feature of relating
quantum to classical properties, but its use for estimating the
spectral correlation functions still raises questions.  The problem is
that the formula is rigorous in general only for times shorter than
the Ehrenfest time: $n \leq \log N$ (and $N \to \infty$).  Yet,
long-time traces ($n \sim N$) are needed to obtain spectral
correlation functions at the scale of the mean level spacing where
universality emerges.  The diagonal approximation, which assumes
statistical independence of the different periodic orbits, is
unsatisfactory at large times where the exponential proliferation of
periodic orbits clashes with the finite ($N$) number of eigenvalues:
the classical information is then overcomplete, which implies some
sort of hidden correlation between the contributions from classical
orbits.  A recipe to overcome this difficulty has been devised by
Bogomolnyi and Keating \cite{bogokeating}, but so far lacks rigorous
justification.

To bypass these problems, a second approach to estimate spectral
correlations has recently emerged, inspired by the study of disordered
metals.  It consists in expressing the correlation function as a
quantum field theory (or functional integral) of the type of a
non--linear sigma model (NL$\sigma$M).  One then tries to analyze the
functional integral by standard field-theoretic methods such as
perturbation expansion, saddle--point analysis and the renormalization
group.  This approach was first applied successfully to systems with
disorder, where the dynamics is governed by a diffusion operator
\cite{efetov}.  The formalism was later extended to the ``ballistic
case'' \cite{agam,khmel}, and quantum correlation functions were put
in relation with the spectrum of the Frobenius-Perron operator
(i.e.~the evolution operator for classical densities).  Although quite
elegant, this approach suffered from several drawbacks.  Among these
are the appearance of unwanted zero modes around the main saddle
point, and the problem of ``mode locking'' \cite{aos}.  Besides, the
results do not exactly agree with the correlations calculated
numerically for the Riemann zeta function (the prototype of a quantum
chaotic spectral determinant) \cite{bogokeating}; nor do they explain
the non--generic spectral correlations featured by systems with
arithmetic symmetries. More recent treatments of the ballistic
NL$\sigma$M have also stressed the need for averaging over a smooth
disorder if one wants to avoid the above problems \cite{taras,mirlin}.

In an attempt to resolve these uncertainties, we have adapted the
latter approach, which had originally been conceived for Hamiltonian
systems, to the case of quantum maps $U_{\phi,N}$.  Our objective was
to prove the universality of the determinant correlation function
$\Omega_U(\gamma)$ (Section \ref{algebra}) upon averaging w.r.t.~a
suitable measure on ${\rm U}(N)$.  This correlation function is easier
to treat than the two--level correlation function, as it does not
require the use of a supersymmetric representation \cite{agam,khmel}
but can be expressed as an ordinary c-number integral over a
finite-dimensional manifold ${\cal M}_N$ (Section \ref{nlsm}).  We
write this integral in the form
\begin{equation}\label{noave}
  \Omega_U(\gamma) = \int_{{\cal M}_N}dQ \e^{-S(\gamma,U,Q)},
\end{equation}
where $S(\gamma,U,\cdot)$ is called the \emph{effective action}.

For the reasons stated, we will consider averages of $\Omega_U$ with
respect to certain probability measures ${\rm d}{\cal P}_N(U)$.  The
averaged correlation function, denoted by $\Bl \Omega_U (\gamma) \Br$,
can still be obtained by integrating the Boltzmann weight given by
an effective action:
\begin{displaymath}
  \Bl\Omega_U(\gamma)\Br = \int_{{\cal M}_N}dQ \e^{-S_{\rm av}(\gamma,Q)}.
\end{displaymath}
To estimate these integrals, we apply the same technique that was used
in \cite{agam}: we expand $S_{\rm av}(\gamma,Q)$ up to quadratic order
around its saddle points $Q_{\rm crit}$, and perform the Gaussian
integrals.  The result obtained in this way,
\begin{equation}\label{expansion}
  \Bl\Omega_U(\gamma)\Br_{\big|{\rm s.p.\ exp.}} = \sum_{Q_{\rm crit}}
  \big\{{\rm Det} \, \delta^2 S_{\rm av}(\gamma, Q_{\rm crit})
  \big\}^{-1/2}\; \e^{-S_{\rm av}(\gamma,Q_{\rm crit})} \;,
\end{equation}
is called the leading--order saddle--point expansion of the integral.

Owing to the absence of a large parameter in front of the action
$S_{\rm av}$, the expansion is a priori not justified mathematically.
A more careful treatment should in principle include perturbative
corrections around each critical point (we actually compute the
expansion up to two loops in a particular case, see Section
\ref{loops}).

We have succeeded in computing the leading--order term for a few
averaging schemes.  For an individual matrix $U_N$ we can actually
reproduce the exact value of the correlation function (\ref{noave}) in
this way (Section \ref{saddleanalyses}).  In Section \ref{semiclas} we
define a ``semiclassical'' averaging scheme, which we think is a good
candidate to obtain universality of correlations \cite{zirnbauer};
unfortunately, in that case we can only compute the contributions from
the two standard saddle points.

In order to test the leading--order saddle--point approximation, we
selected a sequence of statistical ensembles (i.e.~a sequence of
measures ${\rm d}{\cal P}_N(U)$) for which the averaged correlation
function can be computed exactly, and compared the exact result with
the saddle--point approximation for the corresponding effective action.
All these ensembles are ${\rm U}(N)$--rotation invariant, that is, we
first average over all bases of ${\cal H}_N$ (Section
\ref{eigenbasisave}), then possibly over the spectrum of $U_N$
(Sections \ref{rmtensembles}, \ref{cross}).  In most cases, the
saddle--point expansion of these ensembles yields erroneous results.
We still hope that the expansion is better behaved in the case of
local averages, like the semiclassical one.
 
These disappointing results seem to challenge the use of NL$\sigma$M
methods for the study of quantum ballistic systems, unless our
understanding and control of these methods significantly improves.  In
Section \ref{HeatKernel}, we introduce a ${\rm U}(N)$--isotropic local
averaging scheme which we treat by an alternative method;
unfortunately, this scheme does not discriminate between the different
universal behaviours that are expected for chaotic versus integrable
maps.  Nevertheless, we use it in Section \ref{cross} to compute the
correlations along a crossover between the Poisson and CUE
universality classes.

\section{Algebraic manipulation of $\Omega_U$}
\subsection{Fourier decomposition of $\Omega_U$}\label{FourierOmega}

We first remind the reader of some known results concerning the
correlation function $\Omega_U$ \cite{haake,smilansky}.  The spectral
determinant of $U\in {\rm U}(N)$ may be expanded as
\begin{equation}\label{akdecompo}
  {\rm Det}(1 - s U) = \sum_{k=0}^N s^k a_k(U).
\end{equation}
The unitarity of $U$ implies a ``self-inversive'' property for the
secular coefficients \cite{BBL}:
\begin{displaymath}
  a_{N-k}(U) = {\rm Det}(-U) a_k(\bar U) \;.
\end{displaymath}
Each coefficient $a_k$ may be obtained from the traces $\{ t_l = {\rm
  Tr} U^l\}$ by
\begin{displaymath}
  a_k = -\frac{1}{k}(t_k+\sum_{l=1}^{k-1}a_l\:t_{k-l}) =
  \frac{(-1)^k}{k!} \begin{vmatrix} t_1&t_2&t_3&\ldots&t_k\\ 1
    &t_1&t_2&\ldots&t_{k-1}\\ 0 &2 &t_1&\ddots &t_{k-2}\\ 0 &0 &3
    &&\vdots\\ \vdots&\ddots&&&\vdots\\ 
    0&\hdotsfor{2}&{k-1}&t_1\end{vmatrix} \;.
\end{displaymath}
Because this dependence is highly non--linear, the secular
coefficients inherit non--Gaussian distributions in the RMT ensembles
\cite{haake}.  However, to compute the ensemble averages of
$\Omega_U(\gamma)$ one only needs to know their variances, since
\begin{equation}\label{akdecompo2}
  \Omega_U(\gamma) = \sum_{k=0}^N \gamma^{k-N/2} |a_k|^2 =
  \sum_{k=0}^{N/2}(\gamma^{k-N/2}+\gamma^{N/2-k})|a_k|^2 \;.
\end{equation}
For the Poisson and the CUE ensemble of random matrices, these
variances were computed in \cite{haake}, and have the following
large--$N$ asymptotics:
\begin{eqnarray}
  &&\Bl |a_k|^2\Br_{\rm Poisson} = \binom{N}{k} \;, \qquad \Bl |a_k|^2
  \Br_{\rm CUE} = 1 \;, \\ \label{akaverages} &&\Bl \Omega_U(\e^{\mi
    x/N}) \Br_{\rm Poisson} \sim 2^N \;, \quad \Bl\Omega_U(\e^{\mi
    x/N})\Br_{\rm CUE} \sim N\frac{\sin(x/2)}{x/2} \;.
\label{omegaresults} 
\end{eqnarray}
In \cite{smilansky}, a semiclassical estimation of the $|a_k|^2$ was
given for integrable and chaotic quantum maps.  The authors used the
explicit expression in terms of the traces $t_l$, and estimated the
latter by the Gutzwiller trace formula (\ref{gutzwiller}).  They made
a generalized diagonal approximation treating the traces $t_l$ as
statistically independent variables.  To obtain the correlation
function, one has to estimate the $|a_k|^2$ (and hence the $t_k$) up
to times $k\lesssim N/2$, where the Gutzwiller formula is non--rigorous.

\subsection{Representation-theoretic content of $\Omega_U$}
\label{algebra}

We now introduce a more group-theoretic expression for the correlation
function.  Instead of performing the expansion (\ref{akdecompo}), we
will express $\Omega_U(\gamma)$ as a \emph{character} in a certain
irreducible representation of ${\rm U}(2N)$, which is best described
using the physical language of fermions.

Let ${\cal F}_N$ be the Fock space for $N$ types of fermions $f_i^{
  \vphantom{\dagger}}$, $f^\dagger_i$.  In mathematics ${\cal F}_N$ is
known as the spinor representation space of the group ${\rm Spin}
(2N)$.  Then, for any $N\times N$ unitary matrix $U$,
\begin{displaymath}
  {\rm Det}(1-U) = \Tr_{{\cal F}_N} (-1)^{\sum_i f_i^\dagger f_i^{
      \vphantom{\dagger}}} \; \exp \sum_{i,j=1}^N {f^\dagger_i} (\log
  U)_{ij} f_j \;.
\end{displaymath}
The exponential on the right-hand side can be shown to be
well--defined in spite of the multi-valuedness of $\log U$.  To
account for both determinants, we use $2N$ fermions, whose creation
operators are denoted by $f_{+j}^\dagger$ and $f_{-j}^\dagger$ , $j =
1, \dotsc, N$.  The integration over $\phi$ in the integral
(\ref{omegaU}) projects on the subspace ${\cal F}\defi \Ker(F_+
-F_-)$, where $F_\pm = \sum_i f^\dagger_{\pm i}
f^{\vphantom{\dagger}}_{\pm i}$ are the number operators for the two
types of fermion.  The correlation function reads
\begin{equation}\label{trace}
  \Omega_U(\gamma) = \Tr_{\cal F}\ \gamma^{(F_{+} + F_{-} - N)/2} \exp
  \sum_{i,j=1}^N\ (\log U)_{ij} (f^\dagger_{+i} f_{+j}^{\vphantom{
      \dagger}} - f^\dagger_{-j} f_{-i}^{\vphantom{\dagger}} ) \;.
\end{equation}
The operator under the trace belongs to an irreducible representation
$R$ of the group ${\rm U}(2N)$, realized on the space $\cal F$, which
has dimension $\binom{2N}{N}$.  This representation may be defined
through its Lie algebra version: any skew-hermitian $2N\times 2N$
matrix ${\bf X} = \begin{pmatrix} a &b \\ c &d \end{pmatrix}$ is
represented by the operator
\begin{equation}\label{u2N}
  R({\bf X}) = \sum_{i,j=1}^N a_{ij} \, f^\dagger_{+i} f_{+j}^{
    \vphantom{\dagger}} + b_{ij} \, f^\dagger_{+i}f^\dagger_{-j} +
  c_{ij}\,f_{-i} f_{+j} + d_{ij} \, f_{-i}^{\vphantom{\dagger}}
  f^\dagger_{-j} \;.
\end{equation}
By exponentiating, $R(\exp {\bf X}) = \exp R({\bf X})$, we obtain a
${\rm U}(2N)$--representation, which we still denote by $R$.  The
correlation function $\Omega_U(\gamma)$ for any $N\times N$ unitary
matrix $U$ may be recast as a character in this representation:
\begin{eqnarray}\label{character}
    \Omega_U(\gamma) &=& \gamma^{-N/2} {\rm Det}(U)^{-1}\;\Tr\ R({\bf
      \Gamma U})\\ \mbox{where}\ &&{\bf U} \defi \begin{pmatrix} U &0\\ 
      0 &U\end{pmatrix} , \quad {\bf\Gamma}\defi \begin{pmatrix}
      \gamma &0\\0&1\end{pmatrix} \in {\rm U}(2N)\nonumber \;.
\end{eqnarray}
As it stands, the construction assumes $\gamma = \e^{\mi\theta} \in
{\rm U}(1)$.  It can also be used for other values of $\gamma$, since
$R$ naturally extends to a representation of ${\rm GL}(2N,{\mathbb
  C})$.  In the following, matrices in bold print will always be of
size $2N\times 2N$.

The assignment $U \mapsto {\bf U}$ embeds ${\rm U}(N)$ into ${\rm U}
(2N)$.  By this embedding, $R$ restricts to a \emph{reducible}
representation of ${\rm U}(N)$ on $\cal F$, which we simply denote by
$R(U)$.  To express the correlation function, we may also consider the
${\rm U}(N)$--representation $R_{-1}(U) \defi \det(U)^{-1} R(U)$.

In the next section, we decompose $R(U)$ (or equivalently $R_{-1}(U)$)
into irreducible representations (irreps) of ${\rm U}(N)$, thus
expressing the correlation function $\Omega_U(\gamma)$ as a sum of
${\rm U}(N)$--characters.

\subsection{$\Omega_U$ as a sum of ${\rm U}(N)$--characters}
\label{sumofcharacters}

The crucial mathematical tool to use is the \emph{dual pair} structure
\cite{howe}.  The subalgebra $\{ X_N \otimes \mathbb{I}_2 ~|~ X_N \in
\mathfrak{u}(N)\}$ of $\mathfrak{u}(2N)$ commutes with the subalgebra
$\{ \mathbb{I}_N \otimes x_2 ~|~ x_2 \in \mathfrak{u} (2)\}$, and each
is the commutant of the other inside $\mathfrak{u} (2N)$: they are
said to form a dual pair.  This means that for all $U\in {\rm U}(N)$,
the operator $R(U)$ commutes with the set
\begin{eqnarray*}
  J_\uparrow &=& \sum_i f^\dagger_{+i}f^\dagger_{-i}\;, \\ 
  J_\downarrow &=& \sum_i f_{-i}f_{+i} \;, \\ J_0 &=& F_+ + F_- -N \;.
\end{eqnarray*}
The operators $J_0$, $J^\uparrow$ and $J^\downarrow$ generate an
$\mathfrak{su}(2)$ algebra.  The equation $J_0 R(U) = R(U) J_0$
implies that $R(U)$ conserves the total number of particles and hence
acts inside the subspaces ${\cal F}^p = {\cal F} \cap \Ker( F_+ + F_-
- 2p)$.

The dual pair structure provides us with a prescription \cite{howe} to
decompose $R(U)$.  Inside the reduced Fock space ${\cal F}$, we
consider the subspace of lowest ${\rm SU}(2)$ weights, ${}^0{\cal F} =
{\cal F} \cap \Ker J_\downarrow$, and expand it according to its
particle content: ${}^0{\cal F}^p \defi {}^0{\cal F} \cap {\cal
  F}^p$.  Classical results of invariant theory, due mostly to H.~Weyl
\cite{weyl} and succinctly summarized by R.~Howe \cite{howe}, amount
to the following statements:
\begin{itemize}
\item The operator $R(U)$ acts inside each space ${}^0{\cal F}^p$,
  through a certain irrep $\tilde\rho_p(U)$ of ${\rm U}(N)$.
  Equivalently, $R_{-1}(U)$ acts on this space through $\rho_p(U) =
  {\rm Det}(U)^{-1} \tilde \rho_p(U)$.  Furthermore, two irreps
  $\rho_p$ and $\rho_{p'}$ are inequivalent if $p \neq p'$.
\item The image of ${}^0{\cal F}^p$ under $(J_\uparrow)^k$ is the
  space $^k{\cal F}^{p+k}\subset {\cal F}^{p+k}$ which is either
  trivial (if $k>N-2p$) or carries the irrep $\rho_p$ (if $k\leq
  N-2p$).  The operators $J_\uparrow$, $J_\downarrow$, $J_0$ act on
  this tower of spaces according to the $\mathfrak{su}(2)$--irrep of
  dimension $N-2p+1$.
\item The direct sum of these towers exhausts ${\cal F}$.
\end{itemize}
We summarize these statements in the following diagram.  All entries
in a given row are subspaces containing the same number of fermions;
all entries in a given column (or tower) carry the same ${\rm U}
(N)$--irrep.  We only show the case where $N$ is an even integer (the
odd--$N$ case being very similar):
\begin{equation}\label{irrepdecompo}
  \begin{array}{ccccccccc}
    {\cal F}^N&=&^N{\cal F}^N&&&&&&\\ \vspace{.2cm} &&\big\uparrow
    J_\uparrow&&&&&&\\ {\cal F}^{N-1}&=&^{N-1}{\cal
      F}^{N-1}&\oplus&^{N-2}{\cal F}^{N-1}&&&&\\ \vdots&&\big\uparrow
    J_\uparrow&&\big\uparrow J_\uparrow&&\ddots&&\\ 
    \vdots&&\vdots&&\vdots&&&&^0{\cal F}^{N/2}\\ 
    \vspace{.2cm}\vdots&&\big\uparrow J_\uparrow&&\big\uparrow
    J_\uparrow&&\big\uparrow J_\uparrow&&\\ {\cal F}^2&=&^2{\cal
      F}^2&\oplus&^1{\cal F}^2&\oplus&^0{\cal F}^2&&\\ \vspace{.2cm}
    &&\big\uparrow J_\uparrow&&\big\uparrow J_\uparrow&&&&\\ {\cal
      F}^1&=&^1{\cal F}^1&\oplus&^0{\cal F}^1&&&&\\ \vspace{.2cm}
    &&\big\uparrow J_\uparrow&&&&&&\\ {\cal F}^0&=&^0{\cal
      F}^0&&&&&&\\ U(N)-{\rm
      irreps}:&&\rho_0&&\rho_1&&\rho_2&\hdots&\rho_{N/2}
  \end{array}
\end{equation}
The leftmost tower on the right-hand side carries the trivial
$\mathfrak{u}(N)$--irrep, so all spaces $^p{\cal F}^p = (J_\uparrow)^p
\ {}^0{\cal F}^0$ are one-dimensional.
 
Each irrep $\rho_p$ (or $\tilde\rho_p$) may be described by a Young
diagram.  $\rho_p$ mixes the action of $U$ on $p$ fermions $f_+$ with
the action of $\overline{U}$ on $p$ fermions $f_-$.  Owing to
antisymmetrization, it corresponds to the diagram with $p$ rows of
length $2$ followed by $N-2p$ rows of length one:
\begin{displaymath}
  \tilde\rho_p(U_N) = {\rm Det}(U_N) \rho_p(U_N) = U_N^{[2^p\ 
    1^{N-2p}]} \;.
\end{displaymath}

In view of the above diagram, the dimensions of the representation
spaces ${}^k{\cal F}^{k+p}$ follow immediately from those of the
spaces ${\cal F}^p$:
\begin{eqnarray}\label{dimrhop}
  {\rm dim}({}^k{\cal F}^{k+p}) &=& {\rm dim}({\cal F}^p) - {\rm
    dim}({\cal F}^{p-1}) \nonumber \\ &=& \binom{N}{p}^2 -
  \binom{N}{p-1}^2 \;.
\end{eqnarray}
By doing the sum over each $\mathfrak{su}(2)$--multiplet we can now
express the correlation function (\ref{character}) in terms of the
irreps $\rho_p$:
\begin{equation}\label{rhodecompo}
  \Omega_U(\gamma) = \sum_{p=0}^{N/2} \Tr \rho_p(U)\ 
  \frac{\gamma^{p-N/2}-\gamma^{N/2+1-p}}{1-\gamma} \;,
\end{equation}
or, making the substitution $\gamma = \e^{\mi x/N}$, 
\begin{displaymath}
  \Omega_U(\e^{\mi x/N}) = \sum_{p=0}^{N/2} \Tr \rho_p(U)\ {\sin
    \left( {x\over 2} (1-\frac{2p-1}{N}) \right) \over \sin({x \over
      2N})} \;.
\end{displaymath}
For large values of $N$ we may replace the denominator $\sin(x/2N)$ by
$x/2N$.  A quick comparison shows that this decomposition is actually
equivalent to the pedestrian expansion (\ref{akdecompo2}) written down
in Section \ref{FourierOmega}.  The squared coefficients $|a_k|^2$ now
acquire a representation-theoretic meaning:
\begin{equation}\label{akrhop}
  \forall p\leq N/2 : \quad |a_p(U)|^2 = \Tr_{{\cal F}^p}\ R_{-1}(U)
  =\sum_{k=0}^p \Tr \rho_k(U) \;,
\end{equation}
or equivalently,
\begin{displaymath}
  \Tr \ \rho_p(U) = |a_p(U)|^2-|a_{p-1}(U)|^2 \;.
\end{displaymath}

As it stands, the decomposition into irreducibles (\ref{rhodecompo})
is not very informative if one takes for $U$ the matrix of a quantum
map.  We have no way a priori to estimate the character ${\rm Tr}
\rho_p (U_N)$ from semiclassical information, except by using the
relationship, via the $|a_k|^2$, to the original traces ${\rm Tr}
(U_N^k)$, as was done in \cite{smilansky}.  This decomposition will,
however, allow us to obtain rigorous results when adopting a ${\rm
  U}(N)$--isotropic averaging centered around $U_N$ (see Section
\ref{HeatKernel}).

\subsection{$\Omega_U$ as a coherent--state integral}
\label{nlsm}

Instead of decomposing the character $\Tr R_{-1}(U)$ into
irreducibles, we can rewrite it as an integral over the symmetric
space ${\cal M}_N = {\rm U}(2N) / {\rm U}(N) \times {\rm U}(N)$.  This
integral can be interpreted as a variant of the non--linear sigma model
used in \cite{agam} to study the spectral statistics of quantum
chaotic Hamiltonians on infinite-dimensional Hilbert spaces.  In our
case the integral representation is exact, and is well--defined
mathematically.

To write the character $\Omega_U(\gamma)$ as an integral, one uses the
coherent states $R({\bf g})|0\rangle$, where $|0\rangle$ is the vacuum
of ${\cal F}$ and ${\bf g}$ any matrix in ${\rm U}(2N)$.  These
coherent states provide a resolution of unity on ${\cal F}$, i.e. they can be
combined to build the orthogonal projector on ${\cal F}$, as:
\begin{displaymath}
  P_{\cal F}\defi \int\limits_{{\rm U}(2N)} d{\bf g}\ \ R({\bf g}) |0
  \rangle \langle 0|R({\bf g})^{-1} \;,
\end{displaymath}
where the Haar measure $d{\bf g}$ has to be suitably
normalized.  Let $H_N$ be the block-diagonal subgroup ${\rm U}(N)
\times {\rm U}(N)$ of ${\rm U}(2N)$.  Then for all ${\bf h}\in H_N$,
the states $R({\bf g}) | 0 \rangle$ and $R({\bf g} {\bf h})|0\rangle$
only differ by a phase factor.  Therefore, it suffices to integrate
over the equivalence classes in ${\rm U}(2N)$ modulo $H_N$:
\begin{displaymath}
  P_{\cal F} = \int\limits_{{\rm U}(2N)/H_N} d[{\bf g}]_H \ R({\bf g})
  | 0 \rangle \langle 0|R({\bf g})^{-1}.
\end{displaymath}
It is convenient to represent the $H_N$--equivalence classes (i.e.~the
points on ${\cal M}_N$) by $2N \times 2N$ matrices.  To each ${\bf g}
\in {\rm U}(2N)$ one associates $Q_{\bf g} = {\bf g}\Sigma_3 {\bf
  g}^{-1}$, where $\Sigma_3 = \mathbb{I}_N \otimes \sigma_3$.  The set
of all these matrices $Q$ is isomorphic to ${\cal M}_N$.  It is the
set of all Hermitian matrices with two eigenvalues, $+1$ and $-1$,
each with multiplicity $N$.  This non--linear set of matrices is
naturally equipped with ${\rm U}
(2N)$--invariant symplectic structure and metric 
(and therefore an invariant measure $dQ$).

The matrix elements $Q_{ij}$ are not all independent, and for
practical calculations we need to introduce a bona fide coordinate
system on ${\cal M}_N$.  If we denote by $Q_{\bf 12},\ Q_{\bf 22}$ the
two $N \times N$ blocks in the right half of the matrix $Q$, the
entries of the complex matrix $Z = Q_{\bf 12} (Q_{\bf 22} - 1)^{-1}$
are good coordinates on the open subset of ${\cal M}_N$ where $(Q_{\bf
  22}-1)$ is invertible.  Geometrically, these $N \times N$ complex
coordinates represent a certain stereographic mapping of ${\cal M}_N$
onto $\mathbb{C}^{N \times N}$.  The matrix $Z$ corresponding to a
point $Q_{\bf g}$ can be extracted from the Gaussian decomposition of
${\bf g}$:
\begin{equation}\label{gaussian}
  {\bf g} = \begin{pmatrix}1&Z\\0&1\end{pmatrix}
  \begin{pmatrix}A&0\\C&D\end{pmatrix} \;.
\end{equation}

These complex coordinates also provide a simple definition of the
coherent states.  Indeed, $R({\bf g}) | 0 \rangle$ is co-linear with
\begin{equation}\label{cs}
  |Z\rangle \defi \exp\{ \sum_{i,j=1}^N f^\dagger_{+i}
  Z_{ij}^{\vphantom{\dagger}} f^\dagger_{-j} \} |0\rangle = \exp \{ R
  \begin{pmatrix}0&Z\\0&0\end{pmatrix}\} | 0 \rangle 
  = R\begin{pmatrix}1&Z\\0&1\end{pmatrix} | 0 \rangle \;.
\end{equation}
As it stands, $|Z\rangle$ is not normalized, but has the following
properties:
\begin{itemize}
\item The overlap between two coherent states reads $\langle Z |
  Z^\prime \rangle = {\rm Det}(1 + Z^\dagger Z^\prime)$.  In
  particular, the norm of $ | Z \rangle$ is ${\rm Det} (1+Z^\dagger
  Z)^{1/2}$.
\item The resolution of unity takes the form
\begin{equation}
  P_{\cal F}=\int_{\mathbb{C}^{N\times N}} d\mu_N(Z,Z^\dagger)
  \frac{|Z\rangle \langle Z|}{{\rm Det}(1+Z^\dagger Z)}
\end{equation}
where the measure $d\mu_N(Z,Z^\dagger) = C_N \times {\rm Det}
(1+Z^\dagger Z)^{-2N} \prod_{i,j=1}^N d^2 Z_{ij}/\pi$ is the
expression for $dQ$ in the coordinates $Z_{ij}$.  The value of the
normalization factor $C_N$ is given at the end of Appendix B.
\item The group ${\rm U}(2N)$ acts on these coherent states as
  follows: 
\begin{equation}
  R\begin{pmatrix}A&B\\C&D\end{pmatrix}|Z\rangle = {\rm Det} (CZ+D) \;
  | (AZ+B)(CZ+D)^{-1} \rangle \;.
\end{equation}
\end{itemize}
The resolution of unity allows to write the character
(\ref{character}) as
\begin{eqnarray}
  \Omega_U(\gamma) &=& \gamma^{-N/2} {\rm Det}(U)^{-1}\int_{{\cal
      M}_N} d[{\bf g}]_H\ \langle 0|R({\bf g})^{-1} R
  \begin{pmatrix}\gamma U&0\\0&U\end{pmatrix}R({\bf g})|0\rangle 
  \nonumber \\ &=& \gamma^{-N/2} \int\limits_{\mathbb{C}^{N\times N}}
  d\mu_N(Z,Z^\dagger)\ \frac{ {\rm Det}(1+\gamma Z^\dagger
    UZU^{-1})}{{\rm Det} (1 + Z^\dagger Z)} = \int\limits_{{\cal
      M}_N}dQ\ \e^{-S(\gamma,U,Q)} .
  \label{csdecompo}
\end{eqnarray}
This expression is the central result of the current section.  It is
an exact formula, which parallels the ``ballistic'' non--linear sigma
model derived in \cite{agam} for Hamiltonian systems with an
infinite--dimensional Hilbert space.  In our finite-dimensional
framework, the non--local field $Q(q',q)$ of $4\times 4$ supermatrices
on configuration space is replaced by a ``lattice field'' $Q_{i\alpha
  , j\beta}$ of $2\times 2$ matrices (with elements indexed by
$\alpha, \beta$) depending on two discrete positions $i,j$.  The
``effective action'' of the present model
\begin{equation}
  \label{action}
  S(\gamma,U,Q) = - {\rm Tr} \{\log(1+\gamma Z^\dagger UZU^{-1})-
  \log(1+ZZ^\dagger)\}+\frac{N}{2}\log\gamma \;,
\end{equation}
can be presented \cite{aos} in the form 
\begin{eqnarray*}
  S(\e^{\mi x/N},U,Q) &=& - {\rm Tr} \log \big[ \cosh (H_{x,U}) -
  \sinh (H_{x,U}) Q \big] \;,\\ &&{\rm with} \quad H_{x,U}
  \defi\frac{\mi x}{4N}\Sigma_3 +\frac{1}{2}\log {\bf U} \;.
\end{eqnarray*}
In \cite{aos}, this action was further transformed, using the Wigner
representation of wave functions, to obtain the same ballistic
non--linear sigma model as in \cite{agam}.  We will not perform these
steps, which require some further approximations, but rather try to
estimate the integral with the above (purely quantum) effective
action.

\section{Saddle--point analysis of the action $S(\gamma,U,Q)$}
\label{saddleanalyses}

To estimate the field integral of their non--linear sigma model, the
authors in \cite{agam} expand the effective action around two critical
points (usually referred to as saddle points in this context).  Since
there is no large parameter in front of this action, a leading--order
saddle--point expansion -- see Eq.~(\ref{expansion}) -- is a not
justified mathematically a priori.  In the present section we
explicitly compute this expansion for the action (\ref{action}) and
compare it to the results of \cite{agam} and the exact correlation
function.

The saddle points are determined by requiring the variation of the
action to be zero.  In the absence of a large parameter, one first
needs to understand exactly which action to vary.  This point is not
entirely obvious: one might be tempted to lift (part of) the
denominator ${\rm Det}(1+ZZ^\dagger)^{-2N}$ of the measure
$d\mu_N(Z,Z^\dagger)$ into the exponent; this modification of the
effective action would yield a different saddle--point expansion.
However, the requirement of coordinate invariance tells us to keep the
${\rm U}(2N)$--invariant measure $dQ$ as it is, forbidding such
manipulations.  With this convention the saddle--point expansion of
$S(\gamma,U,Q)$ will turn out to yield \emph{the exact
  $\gamma$--dependence} for $\Omega_U(\gamma)$.  In particular, the
problem of ``unphysical zero modes'' occurring in \cite{agam,aos} is
resolved.

We now describe the saddle--point analysis of $S(\gamma,U,Q)$ in some
detail.  We first use the fact that the action is invariant under
simultaneous rotations of both $U$ and $Q$:
\begin{equation}
  \label{covariance}
  S(\gamma,U,Q) = S(\gamma,VUV^{-1},{\bf V}Q{\bf V^{-1}}),
\end{equation}
where we used the shorthand notation ${\bf V} = V\otimes\mathbb{I}_2$,
for $V\in {\rm U}(N)$.  Such a $V$--rotation of $Q$ is an isometry of
the Riemannian manifold ${\cal M}_N$ and leaves the measure $dQ$
invariant.  It therefore suffices to study the simpler situation where
$U$ is \emph{diagonal}: $U\equiv D = {\rm diag} (\e^{\mi \theta_j})$.

One sees from formula (\ref{action}) that the point $Z = 0$ (or
equivalently, $Q = \Sigma_3$) is a saddle point, and the quadratic
approximation to $S$ for small $Z$ reads:
\begin{eqnarray*}
  S(\gamma,D,Z) &\approx& {\textstyle{1 \over 2}} N \log\gamma - {\rm
    Tr}(\gamma Z^\dagger DZD^{-1}-Z^\dagger Z)\\ &\approx& {\textstyle{1
      \over 2}} N \log\gamma+\sum_{i,j=1}^N |Z_{ij}|^2\,
  \left(1-\gamma\e^{\mi(\theta_i-\theta_j)}\right) \;.
\end{eqnarray*}
This saddle point is the only one on ${\cal M}_N$ which is located at
a finite $Z$.  It is sometimes called the ``perturbative'' saddle
point in the physics literature.  For a generic matrix $U$, there are
$N$ directions $Z_{jj}$ that have a coefficient $(1-\gamma)\sim -\mi
x/N$; these directions are called ``zero modes'' \cite{agam}, because
their coefficient vanishes as $x\to 0$.  Doing the integral in this
quadratic approximation around $Z = 0$ yields
\begin{equation}
  \label{zerocontribution}
  \Omega_U(\gamma)_{\big|\Sigma_3} = C_N\ 
  \frac{\gamma^{-N/2}}{(1-\gamma)^N \prod_{i\neq j}
    (1-\gamma\e^{\mi(\theta_i-\theta_j)})} \;.
\end{equation}
We chose to separate the zero mode contributions from the others.

The existence of a second saddle point was pointed out (in the context
of the diffusive non--linear sigma model) in \cite{altshulerandreev}.
It may be exhibited through the change of variable $Z^\prime = 1/Z$,
which amounts to switching to the stereographic projection of ${\cal
  M}_N$ from the antipodal point.  In terms of the new variable
$Z^\prime$, the integrand reads
\begin{displaymath}
  \gamma^{N/2}\ \frac{ {\rm Det}(1+\gamma^{-1}{Z'}^{\dagger} U Z'
    U^{-1})} {{\rm Det}(1+{Z'}^\dagger Z')} \;,
\end{displaymath}
so it has the same structure as the original integrand, but for an
additional prefactor $\gamma^N$ and the replacement $\gamma \to
\gamma^{-1}$ in the determinant.  Quadratic expansion around $Z' = 0$
(or, equivalently, around $Q = -\Sigma_3$) yields
\begin{equation}
  \label{inftycontribution}
  \Omega_U(\gamma)_{\big|-\Sigma_3}= C_N\ 
  \frac{\gamma^{N/2}}{(1-\gamma^{-1})^N\prod_{i\neq j}
    (1-\gamma^{-1}\e^{\mi(\theta_i-\theta_j)})} \;.
\end{equation}
These two saddle points $Q = \pm\Sigma_3$ (we call them ``standard'')
are the only ones taken into account in the treatment of the ballistic
non--linear sigma model in \cite{agam,aos}.  The problem with this
approximation is that, in the limit $\gamma \to 1$, the sum of the two
contributions Eq.~(\ref{zerocontribution}) and
(\ref{inftycontribution}) diverges at least as strongly as
$1/(1-\gamma)^{N-1}$, whereas the exact correlation function is bounded.
This phenomenon was attributed to the $N-1$ ``unphysical'' zero modes
appearing at each saddle point (as opposed to the single ``ergodic''
zero mode $\sum_j Z_{jj}$).  More generally, these contributions
become singular each time $U$ and $\gamma U$ happen to have common
eigenvalues.

We will argue below that this problem with zero modes is actually
resolved by taking into account \emph{further} saddle points of the
effective action.

\subsection{Weyl character formula}\label{weylformula}

To identify all saddle points, we return to the expression
(\ref{csdecompo}) of the integrand.  We still study the case where
$U = D$ is diagonal, and we write ${\bf \Gamma D} \equiv {\rm diag}
(\gamma D,D)$.

Let $\zeta$ be a complex $N \times N$ matrix.  The point $Q_{\bf g}$
of ${\cal M}_N$ is a saddle point of the integrand iff the Taylor
expansion of $\langle\zeta|R({\bf g^{-1}\Gamma Dg})|\zeta\rangle$
around $\zeta = 0$ contains no term linear in $\zeta$ and
$\zeta^\dagger$. (Note that this statement is independent of the
choice of representative ${\bf g}$ for $Q_{\bf g}$.)  Moreover, we do
not want the integrand to vanish at $\zeta = 0$.  If we decompose the
unitary matrix as ${\bf g^{-1}\Gamma Dg} = \begin{pmatrix} a&b \\ c&d
\end{pmatrix}$, these conditions read:
\begin{equation}\label{spcondition}
  b = c = 0 \;, \quad {\rm Det}(d) \neq 0 \;.
\end{equation}
This means that the matrix ${\bf g^{-1}\Gamma Dg}$ (for $\gamma = {\rm
  e}^{{\rm i}x/N} \in {\rm U}(1)$) belongs to the subgroup $H_N$ of
${\rm U}(2N)$, which in turn allows ${\bf g}$ to be written as the
product of a \emph{permutation matrix} ${\bf g_\sigma}$ with some
element ${\bf h}\in H_N$.  By ${\bf g_\sigma}$ we mean the unitary
matrix $({\bf g_\sigma})_{ij}=\delta_{i,\sigma(j)}$, where $\sigma$ is
a permutation of $\{1,\dotsc,2N\}$.  To each permutation $\sigma$
there corresponds a single point $Q_\sigma = {\bf g_\sigma} \Sigma_3
{\bf g_\sigma}^{-1}$.  Moreover, two permutations $\sigma$, $\sigma'$
lead to the same point if $\sigma = \sigma^\prime \tau$ where $\tau$
permutes indices separately inside $\{1,\dotsc,N\}$ and $\{N+1 ,
\dotsc , 2N\}$; this property defines a partition of the symmetric
group $\mathfrak{S}_{2N}$ into $\binom{2N}{N}$ equivalence classes,
each one corresponding to a saddle point of the integrand.  

These
classes are in one-to-one correspondence with the sets $S = \sigma(
\{1,\dotsc,N\})$, so we can write $Q_\sigma=Q_S$. 
$Q_S$ is then the diagonal matrix with entries $+1$ at the positions
$j\in S$, and $-1$ at the positions $j\in \bar S$ (the complement of
$S$ in $\{1,\dotsc,2N\}$).  We partition the set $S$ into $S_1 = S\cap
\{1, \dotsc, N\}$ and $\tilde S_2 = S\cap\{N+1, \dotsc, 2N\}$.  In the
following we will also use the set $S_2 = \{j-N|j\in\tilde S_2\}$, and
the sets $\bar S_1$ and $\bar S_2$ which are the complements in $\{ 1,
\ldots, N \}$ of $S_1$ resp.~$S_2$.  The point $Q_S$ corresponds to
the following (coherent) state in $\cal F$:
\begin{equation}\label{tusigma}
  | S \rangle\defi R({\bf g_\sigma})|0\rangle = \pm \prod_{i\in
    \bar S_1} f^\dagger_{+i}\!  \prod_{j\in S_2}f^\dagger_{-j}
  |0\rangle \;.
\end{equation}
The matrix ${\bf g_\sigma}$ admits a Gaussian decomposition
(\ref{gaussian}) iff $\sigma$ is in the trivial class, i.e.~$S =
\{1,\dotsc,N\}$, which explains why only the perturbative saddle point
$Q = \Sigma_3$ could be exhibited from the $Z$--coordinates.

We now compute the leading--order contribution from each saddle point
$Q_S$.  In the vicinity of $Q_S$ the integrand in (\ref{csdecompo})
takes the values
\begin{displaymath}
  \langle\zeta | R({\bf g}_\sigma^{-1} {\bf \Gamma} {\bf D} {\bf
    g}_\sigma) | \zeta\rangle / \langle\zeta|\zeta\rangle \;,
\end{displaymath}
where the entries of the matrix $\zeta$ are ``small'' ($\zeta$ defines
a local coordinate system near $Q_S$).  We partition the diagonal
matrix ${\bf g_\sigma^{-1}\bf \Gamma D\bf g_\sigma}$ into two halves:
${\bf g_\sigma^{-1}\bf \Gamma D\bf g_\sigma}= {\rm diag} (\Delta_1,
\Delta_2)$.  The above integrand then reads
\begin{displaymath}
  {\rm Det}(D) \; \times \; \frac{{\rm Det}(1+\zeta^\dagger \Delta_1
    \zeta \Delta_2^{-1})} {{\rm Det}(1+\zeta^\dagger\zeta)} \;.
\end{displaymath}
Expanding to quadratic order and integrating over $\zeta, \zeta^
\dagger$, we obtain from the saddle point $Q_S$ a contribution
similar to (\ref{zerocontribution}):
\begin{eqnarray}\label{sigmacontribution}
  \Omega_D(\gamma)_{\big|Q_S} &=& C_N \gamma^{-N/2} \; \prod_{{i
      \in \bar S_1} \atop {j \in S_2}} \gamma {\rm e}^{{\rm i}
    (\theta_i - \theta_j)} \prod_{{i \in S_1} \atop {j \in \bar S_2}}
  \left( 1 - \gamma {\rm e}^{{\rm i} (\theta_i - \theta_j)}
  \right)^{-1} \times \\ &\times& \prod_{{i \in \bar S_1}
    \atop {j \in S_2}} \left( 1 - \gamma^{-1} {\rm e}^{-{\rm i}
      (\theta_i - \theta_j)} \right)^{-1} \prod_{{i \in S_1} \atop {j
      \in \bar S_1}} \left( 1 - {\rm e}^{{\rm i} (\theta_i -
      \theta_j)} \right)^{-1} \prod_{{i \in \bar S_2} \atop {j \in
      S_2}} \left( 1 - {\rm e}^{-{\rm i} (\theta_i - \theta_j)}
  \right)^{-1} \;. \nonumber
\end{eqnarray}
Note that the product contains a factor $(1-\gamma)^{-N+2r}$, with $r
= \sharp(S_1\cap S_2)$.  The most singular case $r = 0$ arises for
$S_2 = \bar S_1$, i.e.~saddle points of the type $\prod_{i\in
  \bar S_1}f^\dagger_{+i} f^\dagger_{-i}|0\rangle$.

In the general case $U = V D V^{-1}$, the saddle points are the points
$Q_{V,S} = {\bf V}Q_S{\bf V^{-1}}$, and they lead to the same
contributions (cf.~the covariance of the action and the measure $dQ$).
The two standard saddle points $Q = \pm\Sigma_3$ are the only ones
unaffected by these $V$--rotations.

It is illuminating to present the result of the approximation
(\ref{sigmacontribution}) in an alternative fashion.  For that
purpose, we denote the non--zero elements of the diagonal matrix ${\bf
  \Gamma} {\bf D}$ by ${\rm e}^{{\rm i}\phi_\nu}$ $(\nu = 1, \ldots ,
2N)$.  The sum of contributions (\ref{sigmacontribution}) can then be
rewritten in the form
\begin{equation}\label{weyldecom}
  \Omega_D(\gamma) = C_N \gamma^{-N/2} \sum_S { \prod_{\mu \in \bar
      S_1} \prod_{\nu \in \tilde S_2} {\rm e}^{{\rm i}(\phi_\mu -
      \phi_\nu)} \over \prod_{\mu \in S} \prod_{\nu \in \bar S} \left(
      1 - {\rm e}^{{\rm i}(\phi_\mu - \phi_\nu)} \right)} \;.
\end{equation}
Save for the prefactor $C_N$, the expression (\ref{weyldecom}) agrees
with the result that follows from the \emph{Weyl character formula}
\cite{knapp} for the trace of $R({\bf \Gamma D})$ over $\cal F$.  In
general, this formula expresses the character of an element of ${\rm
  U}(2N)$ (more generally, ${\rm GL}(2N,\mathbb{C})$) in some
representation $R$ as a sum over all permutations $\sigma \in
\mathfrak{S}_{2N}$ (this being the so-called Weyl group of ${\rm
  U}(2N)$).  In our case, the terms from the $(2N)!$ elements of
$\mathfrak{S}_{2N}$ may be grouped into $\binom{2N}{N}$ classes,
according to the equivalence relation described above.  Since Weyl's
formula is an exact result, the expression (\ref{weyldecom}) remains
finite in the limit $\gamma \to 1$, which means that the singularities
$1/(1-\gamma)^{N-2r}$ of the various terms cancel each other.  The
complete sum over saddle--point contributions thus solves the problem
of ``unphysical zero modes'', i.e.~the divergence problem of the two
standard saddle points.

The mathematical reason behind the ``almost exactness'' of the
leading--order saddle--point expansion is as follows.  The action of
$R({\bf g})$ on coherent states $|Z\rangle$ may be interpreted as the
equivariant action of ${\bf g}$ on the space of holomorphic sections
of a certain complex line bundle ${\cal L}_R$ over ${\cal M}_N$
\cite{stone}.  This equivariant action can be extended to the
(infinite-dimensional) space of square-integrable differential forms
of degree $(0,p)$ on the bundle.  On the enlarged space, the character
becomes a (super)trace, which can still be written as an integral over
${\cal M}_N$.  Owing to an $N = 2$ supersymmetry, the integrand may be
continuously deformed without changing the value of the integral.  In
one limit of the deformation, one gets $\Tr R({\bf g})$; in the other,
the integrand \emph{localizes} at the fixed points of ${\bf g}$ on
${\cal M}_N$, yielding Gaussian integrals around these points.

It turns out that these fixed points coincide with our $Q_{V,S}$, and
their (Gaussian) contributions are equal to (\ref{sigmacontribution}),
save for the prefactor $C_N$.  As a result, the leading--order
saddle--point approximation (for our non--localized integrand)
delivers the correct answer (omitting the prefactor).  In Section
\ref{loops}, we investigate the higher--order terms of the expansion
at $Q=\Sigma_3$ up to two loops: we find that these terms only
renormalize the prefactor $C_N$, without affecting the $U$-- or
$\gamma$--dependence.  We speculate that the (adequately resummed)
full series yields the exact answer, including the correct
normalization.

To achieve agreement with the Weyl character formula, it was crucial
to regard the denominator ${\rm Det}(1+Z^\dagger Z)^{-2N}$ as part of
the measure (as opposed to lifting it into the action).  Indeed, in
order for the mechanism of equivariant localization to take effect,
the integration measure must be ${\rm U}(2N)$--invariant -- a property
not enjoyed by the flat measure $\prod_{i,j} d^2 Z_{ij}$ without the
factor ${\rm Det}(1+Z^\dagger Z)^{-2N}$.

\section{Why do we need averaging?}
\label{sectaverage}

While the Weyl character formula for $\Omega_{U}(\gamma)$ constitutes
an exact result, it is of no use -- at least not as it stands --
towards our goal of proving universality of the correlation function.
This formula relies on the knowledge of the eigenphases
$\e^{\mi\theta_i}$ of $U$, which are not given a priori.  It does not
exhibit the semiclassical features of the quantum map at all.  On the
contrary, it is a ``purely quantum'' decomposition of the correlation
function, a complicated reordering of the Fourier decomposition
(\ref{akdecompo2}).

As was explained in the introduction, it is not conceivable in general
that a universal result for $\Omega_U$ can be obtained without doing
some kind of averaging over the matrix $U$.  Given the results of the
previous section, one might try to perform the averaging term by term
in the Weyl decomposition, hoping that most of the terms might average
to zero.  Such a hope is quickly discouraged by a look at the
expression (\ref{zerocontribution}): aside from having an $N^{\rm
  th}$--order singularity at $\gamma = 1$, whose degree increases each
time some $\gamma \e^{\mi (\theta_i - \theta_j)}$ crosses unity, this
contribution to $\Omega_U(\gamma)$ is \emph{strictly positive} for
real $\gamma < 1$.  We know that the singularities are artifacts of
the Weyl decomposition, as the correlation function $\Omega_U(\gamma)$
itself is uniformly bounded w.r.t.~$U$ and $\gamma$.  Unfortunately,
because of the positivity of (\ref{zerocontribution}) the
singularities can only be removed by reorganizing the entire sum of
contributions, not by averaging individual terms.

For this reason, we will adopt a different strategy: we first perform
the ${\rm d}{\cal P}_N(U)$ average on the \emph{integrand} of the
coherent--state integral, obtaining a new effective action
\begin{equation}\label{averageintegrand}
  \e^{-S_{\rm av}(\gamma,Q)} \defi \Bl \e^{-S(\gamma,U,Q)} \Br_{{\cal
      P}_N}\;.
\end{equation}
We then estimate the resulting $Q$--integral by performing a
saddle--point approximation on the action $S_{\rm av}(Q)$
\cite{zirnbauer}.

A priori, this approximation is no more justified than the one in the
previous section, as $S_{\rm av}$ is preceded by no large parameter
either.  The absence of a large parameter also implies that averaging
and making the saddle--point approximation are non--commuting
operations.  Therefore, the saddle--point expansion of $S_{\rm av}$
will yield qualitatively different results from the direct expansion
for $S(\gamma,U,Q)$.  We explained above that averaging the Weyl
character formula is hopeless for our aims.  The other way around
(i.e.~performing the expansion after averaging the action) will prove
more interesting.

\subsection{Where are the critical points of $S_{\rm av}$?} 
\label{where?}

For any averaging measure ${\rm d}{\cal P}_N$, the two points $Q = \pm
\Sigma_3$ remain saddle points of $S_{\rm av}(\gamma,Q)$.  In the
vicinity of $\Sigma_3$, the integrand expands as
\begin{displaymath}
  \left\langle \frac{{\rm Det}(1+\gamma Z^\dagger UZU^{-1})} {{\rm
        Det} (1+Z^\dagger Z)} \right\rangle \approx \exp {\rm Tr}
  \left( \Bl \gamma Z^\dagger \Ad U \cdot Z \Br - Z^\dagger Z\right) =
  {\rm e}^{-{\rm Tr} \, Z^\dagger ({\mathbb I}- \gamma \langle \Ad
    U \rangle ) Z}
\end{displaymath}
where $\Ad U \cdot Z \defi UZU^{-1}$ is the adjoint action of $U$ on
$Z$.  The approximation is valid for $Z$ small.  For larger values of
$Z$, one should add higher cumulants to the right--hand side.
However, for the time being we stick to the purely quadratic
approximation, and carry out the Gaussian integral to obtain
\begin{equation}\label{Sigma_3}
  \Bl \Omega_U(\gamma) \Br_{\big|\Sigma_3} = C_N \gamma^{-N/2}
  \Det\left(\mathbb{I}-\gamma\Bl \Ad U\Br \right)^{-1} \;.
\end{equation}
When the averaging is absent (that is, ${\rm d}{\cal P}_N$ is a Dirac
$\delta$--measure at $U$), we recover the contribution
(\ref{zerocontribution}).  The saddle point $Q = -\Sigma_3$ yields the
same result, with $\gamma \to \gamma^{-1}$.  On setting $\gamma = \e^{
  \mi x/N}$, the sum of contributions becomes
\begin{equation}\label{+-Sigma_3}
  \Bl\Omega_U(\e^{\mi x/N})\Br_{\big|\Sigma_3\cup-\Sigma_3} = 2 C_N
  \Re \left( \frac{\e^{-\mi x/2}} {\Det\left( \mathbb{I} - \e^{\mi
          x/N} \Bl \Ad U \Br\right)}\right) \;.
\end{equation}
In the next section, we examine the possible occurrence of further
saddle points of $S_{\rm av}$.

\subsubsection{Searching for other saddle points}
\label{qualit}

In Section \ref{weylformula} we located the saddle points of the
function $Q\mapsto \langle Z(Q) | R({\bf \Gamma U})| Z(Q) \rangle$,
using the action of the group ${\rm U}(2N)$ on the coherent states $|
Z \rangle$.  This function may be interpreted as the \emph{Husimi
  function} (or Q-symbol) of the operator $R({\bf \Gamma U})$ acting
on ${\cal F}$, and we denote it by $H_{R({\bf \Gamma U})}(Q)$.  By the
same procedure we can obtain the saddle points of $H_{R({\bf g})}(Q)$
for any non--degenerate matrix ${\bf g}\in {\rm U}(2N)$; in that case,
the saddle points $Q_{\rm crit}$ are given in general by finite
matrices $Z_{\rm crit}^{\vphantom{\dagger}}$ and $Z_{\rm crit}
^\dagger$, which are solutions of the saddle--point equations
\begin{equation}\label{spequations}
  \frac{\partial}{\partial Z_{ij}}H_{R({\bf g})}(Z,Z^\dagger) = 0 =
  \frac{\partial}{\partial \bar Z_{ij}} H_{R({\bf g})} (Z,Z^\dagger)
  \quad (i,j = 1, \ldots, N) \;.
\end{equation}
It is useful to extend $H_{R({\bf g})}$ to a function of two
independent complex matrices $Z, Z^*$ (that makes $2N^2$ complex
variables).  The saddle--point equations pose $2N^2$ constraints on the
degrees of freedom $Z$ and $Z^*$, which yields isolated solutions
$(Z_i, Z^*_i)$, provided that the constraints are independent of each
other.

The \emph{reality} of these solutions (i.e.~$Z_i^*=(Z_i)^\dagger$) is
due to a symmetry of the operator $R({\bf g})$, which is not conserved
if we replace $R({\bf g})$ by any operator ${\cal R}$ on ${\cal F}$.
For instance, if the representation $R$ is extended to matrices ${\bf
  G}\in {\rm GL}(2N,\mathbb{C})$, one can show that the saddle points
of $H_{R({\bf G})}(Z,Z^*)$ are real iff ${\bf G}$ is a normal matrix
(i.e.~${\bf GG^\dagger}={\bf G^\dagger G}$).  We are presently unable
to determine the conditions for the saddle points to be real for the
most general ${\cal R}$.  In any case, the saddle points will be real
if ${\cal R}$ is a Hermitian operator.  The Husimi function is then
real, and Morse theory applies to it.  By Morse's theorem \cite{milnor},
the number of saddle points (which we assume to be isolated) is at
least the sum of all Betti numbers of ${\cal M}_N$, which is
$\binom{2N}{N}$ \cite{leray}.  This is exactly the number of saddle
points we found for $H_{R({\bf X})}(Q)$ when ${\bf X}$ is a $2N\times
2N$ Hermitian matrix, so this function is what is called a
\emph{perfect Morse function} for ${\cal M}_N$.  ${\bf X}$ can be
joined to ${\bf g}\in {\rm U}(2N)$ by a continuous path inside the set
of non--degenerate normal matrices: this explains why $H_{R({\bf \Gamma
    U})}$, although a complex function, still has $\binom{2N}{N}$ real
saddle points.

Unlike reality, the property that the solutions of (\ref{spequations})
are isolated points is robust; $(Z_i,Z^*_i)$ are the common zeros of
$2N^2$ polynomials in $Z$ and $Z^*$, so they are \emph{stable}
w.r.t.~perturbations of the coefficients, as long as the equations do
not become degenerate.  In Section \ref{weylformula} the saddle points
of $H_{R({\bf \Gamma U})}(Q)$ were called $Q_{V,S}$.  We now switch to
such complex coordinates $\zeta$ that a saddle point $Q_{V,S}$ is
situated at $\zeta = 0 = \zeta^\dagger$, and perturb $R({\bf \Gamma
  U})$ in ${\rm GL}({\cal F})$ to ${\cal R} = R({\bf \Gamma U}) + \eps
\, \delta {\cal R}$.  Then for $\eps$ small, $H_{\cal R} (\zeta,
\zeta^ \dagger)$ will have an isolated saddle point at $(\zeta_\eps,
\zeta^* _\eps)$, where both $\zeta_\eps$ and $\zeta^*_\eps$ are of
order $\eps$.  Even if it is not real, this saddle point will
contribute to the integral over ${\cal M}_N$: starting from real
coordinates $\Re\zeta_{ij}, \Im \zeta_{ij}$, we can locally deform the
contour so as to reach the point
\begin{displaymath}
  (\Re\zeta_{ij})^{\rm crit} = (\zeta_{\eps,ij}+\zeta^*_{\eps,ji})/2,
  \qquad (\Im\zeta_{ij})^{\rm crit} = (\zeta_{\eps,ij} -
  \zeta^*_{\eps, ji})/2\mi,
\end{displaymath}
and we can compute the saddle--point expansion of $\int H_{\cal R}
(\Re\zeta, \Im\zeta)$ around it.

The averaged integrands we want to consider are all of the type
$H_{\cal R}(Q)$, where
\begin{displaymath}
  {\cal R} = \int_{{\rm U}(N)} {\rm d}{\cal P}_N(V)\ \frac{1}{{\rm
      Det} V} \; R({\bf \Gamma V}) \;,
\end{displaymath}
and ${\rm d}{\cal P}_N(V)$ is a normalized measure on ${\rm U}(N)$.
If this measure is very strongly peaked near a matrix $U_N$, the
resulting operator will be a perturbation of $R({\bf \Gamma U_N}) /
{\rm Det}U_N$, so the above stability arguments apply: the saddle
points are then isolated points near the unperturbed ones, and they
are ``almost real'' and hence will lie on the integration contour
after a slight contour deformation.

For less concentrated measures ${\rm d}{\cal P}_N(V)$, the structure
of the saddle points can change.  In Section \ref{eigenbasisave} we
exhibit an averaging scheme for which the saddle points are real but
not isolated: they form submanifolds of ${\cal M}_N$; this is also the
case for $H_{R({\bf g})}$ if ${\bf g}$ is degenerate.  We do not have
a good estimate of the typical ``width'' of the measure ${\rm d}{\cal
  P}_N(V)$ above which saddle points can coalesce, spread over
higher-dimensional sets, or cease to contribute to the integral (for
instance when they depart too far away from reality).

In general, we are unable to explicitly locate these extra saddle
points, even for the relatively narrow averages described in Sections
\ref{semiclas} and \ref{HeatKernel}; consequently, we cannot do better
than stick to the approximation (\ref{+-Sigma_3}) to describe the
correlation function.  The remaining task then is to investigate the
spectrum of the operator $\Bl \Ad U \Br$, which depends on $U$ and on
${\rm d}{\cal P}_N$.

\subsection{Common spectral features of $\Bl\Ad U\Br$}

The spectrum of $\Bl \Ad U \Br$ has a few features that are
independent of the averaging scheme.  Before averaging, the eigenvalue
unity occurs in $\Ad U$ with multiplicity $N$, corresponding to the
$N$--dimensional space spanned by the $U$--eigenstate projectors $|
\psi_j \rangle \langle \psi_j|$ $(j = 1, \ldots, N)$, and the
remaining $N^2-N$ eigenvalues lie on the unit circle.  After
averaging, only the uniform mode $\mathbb{I}_N = \sum_j |\psi_j\rangle
\langle \psi_j|$ is left with eigenvalue at unity, while all other
eigenvalues have moved inside the unit disk.  As a result, the sum of
the contributions (\ref{+-Sigma_3}) stays finite in the limit $\gamma
\to 1$.  Averaging thus removes the ``unphysical zero mode'' problem
associated with the two standard saddle points in Section
\ref{saddleanalyses}.

More precisely, the large--$N$ behaviour of $\Bl \Omega_U(\e^{\mi
  x/N})\Br_{|\pm\Sigma_3}$ for finite $x$ mostly depends on the positions of
the eigenvalues of $\Bl
\Ad U \Br$ \emph{closest to unity}.  
Within the approximation (\ref{+-Sigma_3}), these
eigenvalues are \emph{the relevant dynamical data} of the correlation
function.

\subsection{Semiclassical averaging}\label{semiclas}

In \cite{zirnbauer}, a semiclassical averaging scheme around a
quantized map $U_{\phi,N}$ was proposed as a promising candidate to
obtain universal spectral statistics, differentiating between
integrability vs.~chaotic behaviour of the classical map $\phi$.  One
chooses a finite set of Hamiltonian functions $H_j$, corresponding to
Hamiltonian vector fields $\Xi_{H_j}$ $(j = 1, \ldots, r)$, on the
classical phase space.  These Hamiltonians are quantized on each of
the quantum Hilbert spaces ${\cal H}_N$, yielding operators $\{\hat
H_j\}$, which are represented by Hermitian $N\times N$ matrices
w.r.t.~an orthonormal basis of ${\cal H}_N$.  An ensemble average is
then introduced by
\begin{enumerate}
\item composing $U_{\phi,N}$ with the operator $\exp \left( - \mi
    \sum_j t_j \hat H_j / \hbar \right)$, where the ``times'' $t_j$
  are real numbers;
\item averaging over the parameters $t_j$ in a window around the
  origin of width $\eps$ using, for instance, the Gaussian weight
  $(\eps^2 \pi)^{- r/2} \e^{ - \sum_j t_j^2 / \eps^2}$.
\end{enumerate}
The width $\eps$ is taken to be $\hbar$--dependent: $\eps \sim
\hbar^\alpha \sim N^{-\alpha}$ for some $1 > \alpha > 0$, so that the
probability measure for the classical maps $\exp \left( \sum_j t_j
  \Xi_{H_j} \right) \circ \phi$ shrinks to a single point,
$\phi$, in the classical limit $N \to \infty$.  The set of
Hamiltonians $\{ H_j \}$ is chosen once and for all, and is
independent of $N$ and the map $\phi$.  The only constraint on this
set is that the second--order differential operator $- \Delta = \sum_j
\Xi_{H_j}^2$ must be \emph{elliptic} \cite{zirnbauer}.

As explained in the introduction, this averaging procedure is
introduced in order to suppress the non--generic spectral statistics of
quantum chaotic systems with arithmetic symmetries.  In this respect
we must mention the results obtained in \cite{mezzadri}, where the
authors show how non--linear perturbations of quantum cat maps exhibit
generic spectral statistics, as long as one perturbs in both
directions of the 2--dimensional phase space; in contradistinction,
perturbation in a single direction may leave one arithmetric symmetry
intact, leading to non--generic quantum spectral statistics.  This need
for ``phase-space-isotropy'' of the perturbations is very similar to
our ellipticity requirement: $\Delta$ is elliptic only if the vector
fields $\Xi_{H_j}$ span the whole tangent space at every point of
phase space.

Some recent articles \cite{haake01,fishman} have dealt with the
spectral analysis of the operator $\Bl \Ad U \Br_{\rm semiclas}$, and
obtained interesting results concerning its largest eigenvalues.  For a
classically chaotic map, these were shown to converge (as $N \to
\infty$) to the \emph{Ruelle-Pollicott resonances} of the corresponding
Frobenius--Perron operator \cite{ruelle}.  These resonances are inside
the unit circle, which means that $\Bl \Ad U_N \Br_{\rm semiclas}$ has
a \emph{finite gap} between unity and the rest of the spectrum, for $N
\to \infty$.  The huge majority of eigenvalues tend to accumulate on
the origin.

These properties allow us to estimate the contribution from the two
standard saddle points for the case of a quantum chaotic map.  To
lowest order in $1/N$,
\begin{equation}\label{zirn}
  \Bl\Omega_{U_N}(\gamma = \e^{\mi x/N})\Br_{{\rm semiclas} \big
    |\pm\Sigma_3} \stackrel{N\to\infty}{\approx} \frac{N\,C_N}
  {\Det_\bot(\mathbb{I}-\Bl \Ad U \Br_{{\rm semiclas}})}\;
  \frac{\sin(x/2)}{x/2}
\end{equation}
where $\Det_\bot$ means that the determinant is computed after
restriction to the traceless matrices, i.e.~to the subspace orthogonal
to the uniform mode $\mathbb{I}_N$.  Apart from the non--universal
prefactor, the $x$--dependence agrees with the CUE result
(\ref{omegaresults}) in the limit of large matrices.
 
In the case of an integrable map, the eigenvalues of $\Bl \Ad U_N
\Br_{\rm semiclas}$ behave differently: some of them populate more and
more densely a few curves which connect the origin to some point on
the unit circle (including unity).  For this reason, one cannot
separate unity in $\Det \left( {\mathbb I} - \e^{\mi x/N} \Bl \Ad U
  \Br_{\rm semiclas} \right)$ from the rest of the spectrum.  All we
can say is that the approximation (\ref{+-Sigma_3}) does not yield the
CUE formula in that case (in general it does not yield the Poisson
answer either).

\subsubsection{Warning} \label{warning}

One might be tempted to present formula (\ref{zirn}) as a
``physicist's proof'' of a weak universality conjecture for quantum
chaotic maps.  The reason why it is not a proof is clear:
\begin{itemize}
\item As was explained in Section \ref{qualit}, there certainly exist
  other saddle points of (\ref{averageintegrand}).  The calculation of
  their contributions is a difficult task, which we have not yet
  performed.  It is far from obvious why these saddle points should be
  less important than $Q = \pm \Sigma_3$ in the semiclassical
  averaging scheme.
\item As was emphasized before, there is no large parameter in front
  of the effective action.  Without such a parameter, the correction
  terms of the asymptotic expansion around each saddle point are not
  small, and their neglect in the formula (\ref{zirn}) seems to be
  unjustified.
\end{itemize}
The second worry is addressed in the next subsection.

\subsection{Loop expansion}
\label{loops}

We are now going to investigate those corrections to the formula
(\ref{zirn}) that result from systematically expanding around the
saddle point $Q = \Sigma_3$.  The computations will be done up to what
is called two--loop order in field--theoretic language.

As a first step, we approximate the integrand by taking the ensemble
average inside the determinant:
\begin{equation}\label{approx}
  \left\langle {\rm Det} \left( 1 + \gamma Z^\dagger {\rm Ad}U \cdot Z
    \right) \right\rangle_{\rm semiclas} \approx {\rm Det} \left( 1 +
    \gamma Z^\dagger \langle {\rm Ad}U \rangle_{\rm semiclas} \cdot Z
  \right) \;.
\end{equation}
Although $\eps$, the ``width'' of the perturbation, decreases like
$\hbar^\alpha$, its effect is strong enough to completely modify the
spectrum of ${\rm Ad}U$, even in the semiclassical limit.  This shows
that the above approximation is not necessarily valid if we just
suppose that the matrices $Z$ are bounded (uniformly w.r.t.~$N$) in
the operator norm on ${\cal H}_N$.  Using the expansion
\begin{displaymath}
  {\rm Det}(1+A) = 1+\Tr A + \sum_{j=2}^N\Tr (\wedge^j A) \;,
\end{displaymath}
Eq.~(\ref{approx}) will hold as long as the terms for $j\ge 2$ are
small compared to $\Tr A$.  A sufficient condition for that is
$\Tr(|A|)\ll 1$, where $|A|\defi \sqrt{A^\dagger A}$.  Upon the
replacement $A = \gamma Z^\dagger {\rm Ad}U \cdot Z$, this condition
will be met if
\begin{equation}
  \label{condition}
  \Tr(Z^\dagger Z) = \sum_{i,j=1}^N|Z_{ij}|^2\ll 1,
\end{equation} 
uniformly w.r.t.~$N$. It would be desirable to better control the
error in (\ref{approx}) for the larger set of matrices $Z$ satisfying
($N$--uniformly) $\|Z\|_{{\cal L (H}_N)}\le \rm{const}$.

Taking (\ref{approx}) for granted, we proceed to the computation of
higher loops.  To simplify the notation we abbreviate $T \defi \gamma
\langle {\rm Ad} U \rangle_{\rm semiclas}$.  Next we formally
introduce a parameter $M$ (which will be reset to unity at the end of
the calculation) by making in the integrand the replacement
\begin{displaymath}
  {{\rm Det}(1 + Z^\dagger T Z) \over {\rm Det}(1 + Z^\dagger Z)} \to
  \left( {{\rm Det}(1 + Z^\dagger T Z) \over {\rm Det}(1 + Z^\dagger
      Z)} \right)^M \;.
\end{displaymath}
A contribution to the perturbative saddle--point expansion is said to
be of $n$--loop order if it varies as $M^{-n}$ relative to the
leading--order term.  On rescaling the integration variables to $\zeta
= Z \sqrt{M}$ and $\zeta^\dagger = Z^\dagger \sqrt{M}$, the $1/M$
expansion of the integrand looks as follows:
\begin{displaymath}
  d\mu_N(Z,Z^\dagger) \, {{\rm Det}^M(1 + Z^\dagger T Z) \over{\rm
      Det}^M (1 + Z^\dagger Z)} = C_N \prod_{i,j = 1}^N {d^2
    \zeta_{ij} \over \pi M} \, {\rm e}^{- {\rm Tr} \zeta^\dagger ( 1 -
    T ) \zeta} \left( 1 + M^{-1} f_1 + M^{-2} f_2 + ... \right)
\end{displaymath}
where $f_1$ and $f_2$ are the one--loop and two--loop terms,
respectively, and are given by
\begin{eqnarray*}
  f_1 &=& {\textstyle{1 \over 2}} {\rm Tr}(\zeta^\dagger \zeta)^2 -
  {\textstyle{1 \over 2}} {\rm Tr}(\zeta^\dagger T\zeta)^2 - 2N {\rm
    Tr} \zeta^\dagger \zeta \;, \\ f_2 &=& - {\textstyle{1 \over 3}}
  {\rm Tr}(\zeta^\dagger \zeta)^3 + {\textstyle{1 \over 3}} {\rm
    Tr}(\zeta^\dagger T\zeta)^3 + {\textstyle{1 \over 8}} \left[ {\rm
      Tr}(\zeta^\dagger \zeta)^2 - {\rm Tr}(\zeta^\dagger T\zeta)^2
  \right]^2 \\ &+& 2N^2 ({\rm Tr}\zeta^\dagger \zeta)^2 + N {\rm
    Tr}(\zeta^\dagger \zeta)^2 - N {\rm Tr}(\zeta^\dagger \zeta)
  \left( {\rm Tr}(\zeta^\dagger \zeta)^2 - {\rm Tr}(\zeta^\dagger
    T\zeta)^2 \right) \;.
\end{eqnarray*}
The Gaussian integral at leading order just yields the result
(\ref{Sigma_3}).  Using standard diagrammatic techniques to do the
one--loop integral we find the following expression:
\begin{eqnarray*}
  &&{\textstyle{1 \over 2}} C_N M^{-N\times N} \, {\rm Det}(1-T)^{-1}
  \Big\{ -4N \sum_{ij} \left( {1 \over 1-T} \right)_{ij,ij} \\ &+&
  \sum_{ijkl} \left( {1 \over 1-T} \right)_{ij,kj} \left( {1 \over
      1-T} \right)_{kl,il} - \sum_{ijkl} \left( {T \over 1-T}
  \right)_{ij,kj} \left( {T \over 1-T} \right)_{kl,il} \\ &+&
  \sum_{ijkl} \left( {1 \over 1-T} \right)_{ij,il} \left( {1 \over
      1-T} \right)_{kl,kj} - \sum_{ijkl} \left( {T \over 1-T}
  \right)_{ij,il} \left( {T \over 1-T} \right)_{kl,kj} \Big\} \;. \\ 
\end{eqnarray*}
By the relation $(1-T)^{-1} = 1 + T(1-T)^{-1}$ these terms combine to
yield the simple answer
\begin{displaymath}
  C_N \int\limits_{{\mathbb C}^{N\times N}} \prod_{i,j=1}^N {d^2
    \zeta_{ij} \over \pi M} \, {\rm e}^{- {\rm Tr} \zeta^\dagger (1 -
    T) \zeta} f_1(\zeta,\zeta^\dagger) = C_N M^{-N\times N} {\rm
    Det}(1-T)^{-1} \left( -N^3 \right) \;.
\end{displaymath}
We see that the dependence of the one-loop contribution on $T$ cancels
completely, leaving only a constant, $-N^3$.  This cancellation is not
accidental but continues to higher loop order.  By a lengthy but
straightforward calculation, the complete perturbative result up to
two--loop order can be shown to be
\begin{eqnarray*}
  &&\int\limits_{{\mathbb C}^{N\times N}} d\mu_N(Z,Z^\dagger) \, {{\rm
      Det}^M (1 + Z^\dagger T Z) \over {\rm Det}^M(1 + Z^\dagger Z)}
  \\ &=& C_N M^{-N\times N} {\rm Det}(1 - T)^{-1} \left( 1 - M^{-1}
    N^3 + M^{-2} ( {\textstyle{1 \over 2}} N^6 + {\textstyle{7 \over
        12}} N^4 - {\textstyle{1 \over 12}} N^2) + {\cal O}(M^{-3})
  \right) \;.
\end{eqnarray*}
Again, all the $T$--dependence has disappeared from the loop
correction terms.  This is true for all $M$ including the case of
interest, $M = 1$.

The cancellation does not come as a total surprise.  The above
perturbation expansion, whose low--order terms we have computed, is
formally identical to the same expansion \emph{before} averaging.  The
latter is obtained from the former by simply substituting $\gamma {\rm
  Ad}U$ for $T = \gamma \langle {\rm Ad} U \rangle_{\rm semiclas}$.
In the case before averaging we know from \cite{stone} that an
index--theoretic mechanism (sometime called
\emph{localization}) causes the perturbation expansion to be
deformable (by an underlying $N = 2$ supersymmetry) to a harmonic
oscillator problem (or, equivalently, a Gaussian integral) at $Z = 0$.
The process of deformation to the Gaussian limit explains why the
dependence on $\gamma {\rm Ad}U$ is exhausted by the leading--order
term.  It leads to the Weyl character formula, which implies that the
contribution to the character from $Z = 0$ (or $Q = \Sigma_3$) is
\emph{exactly} given by
\begin{displaymath}
  \int\limits_{{\mathbb C}^{N\times N}} d\mu_N(Z,Z^\dagger) \, {{\rm
      Det} (1 + Z^\dagger \gamma {\rm Ad}U \cdot Z) \over {\rm Det} (1
    + Z^\dagger Z)} \Big|_{Z = 0,\ {\rm all}\ {\rm orders}}\ =\ {\rm
    Det}({\mathbb I} - \gamma {\rm Ad} U)^{-1}
\end{displaymath}
where the normalization constant $C_N$ has now been replaced by unity.
The last fact provides the raison d'\^etre for the $N$--dependent terms
produced by the loop expansion: their role is to cancel, after proper
resummation, the prefactor $C_N$.  This property does not depend on
the unitarity of $\gamma\Ad U$, so it holds as well after replacing it
by its average.  Thus, after summing all orders of the perturbation
expansion, we expect that the saddle point $Z = 0$ contributes to the
correlation function as
\begin{displaymath}
  \Bl \Omega_U(\gamma) \Br_{\big|\Sigma_3,\ {\rm all}\ {\rm orders}} =
  \gamma^{-N/2}\; \Det \left(\mathbb{I}-\gamma \langle \Ad U \rangle
  \right)^{-1} \;.
\end{displaymath}

This perturbative result should be used with some care.  Although the
function $f(Z,Z^\dagger;T) = \Det(1 + Z^\dagger T Z) / \Det(1 + Z^
\dagger Z)$ is locally well--defined, it does not extend to a global
smooth function on the manifold ${\cal M}_N$ (in particular, this
function is NOT the Husimi function of an operator on ${\cal F}$).
Indeed, setting $Z = z G$ with any invertible matrix $G$ and sending
$z \to \infty$ always leads to the same point $Q = - \Sigma_3$ on
${\cal M}_N$, regardless of which matrix $G$ we choose, whereas the
limit of $f(z G, \bar z G^\dagger;T)$ as $z \to \infty$ does depend on
the choice of $G$.  Thus, the function $f(Z,Z^\dagger;T)$ is not
smooth at $Q = -\Sigma_3$.

This singularity reflects the fact that the cumulants neglected by our
basic approximation (\ref{approx}) are small (compared to the terms
kept) only for small matrices $Z$ (cf. the discussion following
Eq.~(\ref{approx})).  If $Z, Z^\dagger$ (or some matrix elements
thereof) are allowed to go to infinity, the approximation clearly
loses its validity.  To control the error incurred near the saddle
point $Q = -\Sigma_3$, one needs to switch to another scheme, by first
changing coordinates $Z \to 1/Z$ and $Z^\dagger \to 1/Z^\dagger$ and
only afterwards repeating the above steps.  The contribution from $Q =
-\Sigma_3$ can then be calculated in the same way as the one for $Q =
\Sigma_3$.  The treatment of further saddle points remains an open
problem.

What makes this procedure unsatisfactory is that we are simultaneously
working with several approximation schemes, each of which is only
locally controlled.  To localize the integral at the saddle points in
a mathematically rigorous manner, we would need an approximation that
is \emph{globally well--defined} and well--controlled.  It is not clear
whether such an approximation exists, given the stringent requirement
that the integrand should also have the index--theoretic features that
allow localization techniques to be used.

\section{Averaging $U$ over eigenbases}\label{eigenbasisave}

By its definition (\ref{omegaU}) as a correlation function of spectral
determinants, $\Omega_U(\gamma)$ is invariant under any change of
basis $U \mapsto V U V^{-1}$, with $V$ an arbitary unitary matrix.  In
the $Q$--matrix formulation, this invariance is reflected by the
relation $S(\gamma,U,Q) = S(\gamma,VUV^{-1},{\bf V}Q{\bf V}^{-1})$.
Since the transformation $Q \mapsto {\bf V} Q {\bf V}^{-1}$ has unit
Jacobian, we may absorb ${\bf V}$ into the integration variable $Q$
and compute $\Omega_U (\gamma)$ by first averaging ${\rm e}^{- S(
  \gamma,U,Q)}$ over all rotations $U \mapsto V U V^{-1}$:
\begin{equation}\label{Vav}
  \e^{-S_{V{\rm av}}(\gamma,U,Q)}\defi\frac{1}{{\rm Vol} \ {\rm U}(N)}
  \int_{{\rm U}(N)}dV \ \exp \{-S(\gamma,VUV^{-1},Q)\},
\end{equation}
and then integrating ${\rm e}^{-S_{V{\rm av}}}$ over $Q$.  We saw in
Section \ref{weylformula} that if the matrix $D = V^{-1}UV$ is
diagonal, then the saddle points of $S(\gamma,U,Q)$ are situated on
the points $Q_{V,S} = {\bf V}Q_S{\bf V^{-1}}$.  Because the locations
of these points explicitly \emph{depend} on $V$, we expect that a
smoothing mechanism takes place and the divergences of the individual
terms in the Weyl character formula disappear on averaging over $V$.
In fact, as we will see, the expansion obtained by saddle--point
analysis of the effective action $S_{V{\rm av}} (\gamma, U,Q)$ is
qualitatively quite different from Weyl's formula.

\subsection{Analysis around $\pm\Sigma_3$}

We first describe $S_{V{\rm av}}(\gamma,U,Q)$ near the two saddle
points $Q = \pm\Sigma_3$ (cf.~Section \ref{where?}).  The $V$--averaged
adjoint operator $\Bl \Ad \Br_V$ has a rather simple spectrum: unity
is a simple eigenvalue (associated with $\mathbb{I}_N$), and on the
remaining $(N^2-1)$--dimensional space the operator is proportional to
the identity:
\begin{equation}\label{Vavadu}
  \Bl \Ad U\Br_V = P_\mathbb{I} + (1-P_\mathbb{I}) \ \frac{|\Tr
    U|^2-1}{N^2-1} \;.
\end{equation}
($P_\mathbb{I}$ is the orthogonal projector on $\mathbb{I}_N$.)  We
see that $\Bl \Ad U \Br_V$ has a large gap between unity and the
second eigenvalue, and this gap has the maximal degeneracy.  Assuming
that this degenerate eigenvalue is small $(|\Tr U|\ll N)$, we get the
following leading--order contribution: 
\begin{eqnarray}\label{Vinvariantsp}
  \Bl\Omega_U(\e^{\mi x/N})\Br_{V\big |\Sigma_3\cup -\Sigma_3} &\sim&
  \frac{2N\;C_N}{(1-\alpha/N)^{N^2}} \ \frac{\sin\{x(1/2 - \alpha)\}}
  {x},\\ \label{alpha} {\rm with}\quad \alpha &\defi& (|\Tr U|^2-1)/N
  \;.
\end{eqnarray}
Within this approximation, the correlation function depends on $U =
U_{\phi,N}$ only through the simple quantity $|\Tr U|^2$, which can be
estimated semiclassically by the Gutzwiller--Tabor trace formula
(\ref{gutzwiller}): typically, $\alpha$ is of order ${\cal O}(1/N)$
for a chaotic map, and of order ${\cal O}(1)$ for an integrable one.

Notice that, due to the high degeneracy of the second eigenvalue, 
we do not get in general the CUE result (\ref{omegaresults}), 
although this eigenvalue
is far inside the unit circle. This shows that, to obtain
the CUE result (\ref{zirn}), we not only need a finite gap in the spectrum
of $\Bl \Ad U \Bl$,
but also a fast accumulation of the eigenvalues to the origin. The 
precise condition on the eigenvalues is
$\sum_{j=2}^{N^2}\frac{\lambda_j}{1-\lambda_j}\ll N$. In the present
averaging scheme, this means $\alpha\ll 1$.

\subsection{Critical submanifolds}

We need to investigate the possible influence of other saddle points
of $S_{V{\rm av}}(\gamma,U,Q)$; for the present averaging scheme, we 
will explicitly describe a critical set, which we believe to 
be exhaustive.  
The effective action possesses the
symmetry $S_{V{\rm av}}(Q) = S_{V{\rm av}}({\bf W}Q{\bf W^{-1}})$ for
all $W\in {\rm U}(N)$.  Therefore, the saddle points are grouped into
stationary \emph{submanifolds}, each of them invariant under ${\rm
  U}(N)$.

\subsubsection{Description of the manifolds} 

In Appendix A we prove the following statement: for any initial matrix
$U$ and any $\gamma$, the action (\ref{Vav}) is stationary at the
points $Q={\bf W}Q_\sigma{\bf W^{-1}}$, for all rotations $W\in {\rm
  U}(N)$ and any permutation $\sigma$ (see Section \ref{weylformula}
for the notations $\sigma$, $S = S_1\cup S_2$ etc).  Since ${\rm U}
(N)$ is connected, the points ${\cal M}_S \defi \{{\bf W}Q_S{\bf
  W^{-1}}~|~ W \in {\rm U}(N)\}$ form a connected submanifold of
${\cal M}_N$.

Let $\tau$ be any permutation among $N$ indices.  The set $S^\prime =
(\tau(S_1),\tau(S_2))$ is in general different from $S$, and we have
$Q_S \neq Q_{S^\prime}$; however, $Q_{S^\prime}\in{\cal M}_S$, or
equivalently ${\cal M}_S = {\cal M}_{S^\prime}$.  Putting $p = \sharp
S_2$ and $r = \sharp(S_1\cap S_2)$, we find that ${\cal M}_S$ contains
$\binom{N}{p} \binom{p}{r} \binom{N-p}{r}$ different points $Q_{S^
  \prime}$.  The manifolds ${\cal M}_S$ are in one-to-one
correspondence with the integers $(p,r)$, and their total number is
$(N/2+1)^2$ for $N$ even, and $(N+1) (N+3)/4$ for $N$ odd (including
the isolated points $\pm \Sigma_3$ in the count).
 
For generic $U$ and $\gamma\neq 1$ (genericity means here that the
matrix ${\rm diag}(\gamma U,U)$ is not degenerate), we conjecture that
the submanifolds ${\cal M}_{(p,r)}$ exhaust all the critical points of
the action $S_{V{\rm av}}(\gamma,U,Q)$.

\subsubsection{Contributions of the manifolds}

The leading--order contribution of each submanifold ${\cal M}_S$ to the
$Q$--integral is calculated by separating the tangent space at $Q_S$
into two parts, one parallel and another one transverse to ${\cal
  M}_S$.  The integrand in the vicinity of $Q_S$ then reads 
(to quadratic order)
\begin{displaymath}
  \e^{-S_{V{\rm av}}(Q_S)} \e^{-{\rm Hess}_T(X_T) + {\cal
      O}(|X_T|^3)} \;,
\end{displaymath}
where ${\rm Hess}_T$ is the Hessian of $S_{V{\rm av}}$ around ${\cal
  M}_S$, viewed as a non--degenerate quadratic form on the transverse
part of tangent space (coordinatized by $X_T$).  The exact integral
over ${\cal M}_S$ and the Gaussian integral over the
transverse directions yield the contribution
\begin{equation}\label{MScontrib}
  \Bl\Omega_U(\gamma)\Br_{\big|{\cal M}_S} = C_N \ \gamma^{-N/2+p}
  \frac{{\rm Vol} {\cal M}_{S}} {\sqrt{{\rm Det}({\rm Hess}_T)}}
  \e^{-S_{V{\rm av}} (\gamma,Q_S)}.
\end{equation}
In Appendix B, we explicitly compute the volumes of the submanifolds
${\cal M}_S = {\cal M}_{(p,r)}$:
\begin{displaymath}
  {\rm Vol}{\cal M}_{(p,r)} = \frac{(\Gamma(1) \cdots \Gamma(r))^2 \;
    \Gamma(1)\cdots\Gamma(p-r) \; \Gamma(1) \cdots \Gamma(N-p-r)}
  {\Gamma(1) \cdots \Gamma(N)}.
\end{displaymath}
For all submanifolds ${\cal M}_S \neq \{ \pm\Sigma_3 \}$ (i.e.~$0 < p
< N$), these volumes are $N$--exponentially small.  The quantities
${\rm Hess}_T$ and $S_{V{\rm av}}(Q_S)$ depend on $U$ and $\gamma$; we
are unable to compute them in general.  What we know for sure is that
$|\e^{-S_{V{\rm av}}}| \le 1$, since $\e^{-S(\gamma, U,Q)}$ has this
property.  

For a non--degenerate $U$ and $\gamma = \e^{\mi x/N}$, the Hessian
around ${\cal M}_S$ will possess a single eigenvalue that vanishes
with $x$, while all other eigenvalues stay at least of order ${\cal O}
(1)$.  This means that the contribution from ${\cal M}_S$ goes like
$1/x$ as $x\to 0$.  However, the ``particle-hole duality'' between the
submanifolds ${\cal M}_{(p,r)}$ and ${\cal M}_{(N-p,r)}$ cancels this
divergence in the sum of their two contributions (as it does for $\pm
\Sigma_3$).

As a result, we conjecture that each contribution $\Bl \Omega_U
(\gamma) \Br_{\big| {\cal M}_S \cup {\cal M}_{\bar S}}$ is
$x$--uniformly, $N$--exponentially small compared to that 
from $\pm\Sigma_3$ for large
$N$, owing to the small volumes of ${\cal M}_{(p,r)}$.  Since the
number of critical submanifolds grows like $N^2$, we deduce that the
leading--order saddle--point expansion for the action $S_{V{\rm av}}
(U,\gamma,Q)$ can be truncated to (\ref{Vinvariantsp}) for large $N$.

\subsection{Averaging over Random Matrix ensembles}
\label{rmtensembles}

We may go further and average $\e^{-S}$ not only over the conjugates
of a fixed matrix $U$, but also over the spectrum $\{\e^{\mi\theta_j}
\}$.  For instance we can average $U$ over all matrices in ${\rm U}
(N)$, with a weight corresponding to one of the standard random matrix
ensembles (Poisson, ${\rm CUE}$).  The averaged action will be ${\rm
  U}(N)$--rotation invariant, and its saddle points will still lie on
the submanifolds ${\cal M}_{S}$.  As a result, the leading--order
saddle--point (l.o.s.p.) expansion for such ensemble-averaged actions
can again be truncated to the contribution (\ref{Vinvariantsp}), upon
replacing the coefficient $\alpha$ by its average $\Bl\alpha\Br_{\rm
  ensemble}$ over the ensemble considered.

\subsection{Conclusion: no l.o.s.p.~expansion for the $V$--averaged 
  actions}\label{V-results}

The contribution (\ref{Vinvariantsp}) depends in a very simple manner
on the matrix $U$, namely only on its first trace.  This is in
contradiction with the fact that a priori, all traces up to $\Tr(
U^{N/2})$ enter into $\Omega_U(\gamma)$ (cf.~Eq.~(\ref{rhodecompo})).
By selecting some particular cases, it becomes obvious that the
l.o.s.p.~expansion (\ref{Vinvariantsp}) deviates strongly from the
exact correlation function.  The most immediate counterexample is the
Poisson ensemble, whose correlation function is given in
Eq.~(\ref{akaverages}).  For this ensemble, $\Bl\alpha\Br_{\rm
  Poisson} = 1$, which yields the ${\rm CUE}$ result (!) when inserted
into the formula (\ref{Vinvariantsp}).  We are hence forced to abandon
the l.o.s.p.~expansion for the $V$--averaged actions.  

Nevertheless, we hope that this expansion is still meaningful when the
averaging over $U$ is \emph{local} in ${\rm U}(N)$, which is the case
for the semiclassical average in Section \ref{semiclas} (but not for
the $V$--average).  Hopefully, a local average will still conserve
some memory of the ``localization'' property, which entailed the
``almost exactness'' of the l.o.s.p.~expansion for $S(\gamma,U,Q)$.

In the next section, we will consider a local averaging scheme
different from the semiclassical one.  It possesses group--theoretic
properties, which will allow us to analyse it from the character
decomposition (\ref{rhodecompo}) instead of the coherent--state
integral.

\section{Isotropic averaging}\label{HeatKernel}

Starting from a fixed matrix $U$, one may define an \emph{isotropic}
averaging around $U$, by composing $U$ with the $N\times N$ unitary
matrices $\e^{-\mi H}$, weighted by $\exp(-\Tr H^2/4\eps) dH$ with
small $\eps$ (so that the weight is concentrated at the identity).
Isotropy here means that the measure $dH$ is ${\rm U}(N)$--invariant.
Note that this in sharp contrast with the semiclassical averaging of
Section \ref{semiclas}, where $H$ was a linear combination of $f$
matrices $\hat H_j$, with $f$ independent of $N$.  In the
semiclassical case, the perturbation spanned only a $f$--dimensional
submanifold, whereas in the present case the perturbation completely
fills the $N^2$--dimensional $\eps$--ball centered at $H = 0$.

One can replace the Gaussian weight by any positive normalized ${\rm
  U}(N)$--invariant function of $H$.  For our purposes, it is
convenient to use the \emph{heat kernel} on ${\rm U}(N)$, i.e.~the
kernel of the regularizing operator $\exp(- \eps \Delta)$, where
$\Delta$ is the (positive) Laplace-Beltrami operator on ${\rm U}(N)$.
The heat kernel centered on $U$ is defined as follows: 
\begin{eqnarray*}
  \forall\eps>0 ~:~ -\Delta_{V}K_\eps(V,U) &=& \frac{\partial}
  {\partial\eps} K_\eps(V,U) \\ \lim_{\eps\to +0} K_\eps(V,U) &=&
  \delta_U(V) \;.
\end{eqnarray*}
Owing to the compactness of ${\rm U}(N)$, the density $K_\eps(\cdot,
U)$ for any matrix $U$ converges to the uniform density on ${\rm
  U}(N)$ as $\eps\to\infty$.  Switching $\eps$ from $0$ to $\infty$
therefore realizes a crossover from the Dirac delta measure $\delta_U(
\cdot )$ to the Haar (or ${\rm CUE}$) measure.  For small values of
$\eps$, the kernel $K_\eps(V,U) = k_\eps(VU^{-1})$ is concentrated
around $\e^{-\mi H} = VU^{-1}\approx 1$ and is approximately given by
the Gaussian weight introduced above: $k_\eps(\e^{-\mi H}) \sim \exp(
-\Tr H^2/4\eps)$.

Schur's lemma ensures that $\Delta$ is proportional to the identity on
each ${\rm U}(N)$--irreducible subspace of $L^2({\rm U}(N))$.  As a
consequence, its action on each representation matrix $\rho_p(U)$ of
Eq.~(\ref{rhodecompo}) is simply multiplication by a positive factor,
called the quadratic Casimir invariant, which we denote by $\rho_p
(\Delta)$.  In formulas,
\begin{displaymath}
  \int_{{\rm U}(N)} dV \, \rho_p(V) \, K_{\eps}(V,U) \defi \e^{- \eps
    \Delta_U} \rho_p(U) = \e^{-\eps \rho_p(\Delta)} \rho_p(U) \;.
\end{displaymath}
The factor $\rho_p(\Delta)$ may be computed from the Young diagram of
$\rho_p$; a more direct way is to express $\Delta$ in terms of
fermionic operators acting on the Fock space ${\cal F}$: 
\begin{eqnarray*}
  \Delta\big|_{\cal F} &=& \sum_{i,j=1}^N (f^\dagger_{+i} f_{+j}^{
    \vphantom{\dagger}}-f^\dagger_{-j} f_{-i}^{\vphantom{\dagger}})
  (f^\dagger_{+j} f_{+i}^{\vphantom{\dagger}} - f^\dagger_{-i}
  f_{-j}^{\vphantom{\dagger}}) \\ &=& (N+1) (F_{+}+F_{-}) -
  (F_{+}^2+F_{-}^2) - 2J_\uparrow J_\downarrow \;.
\end{eqnarray*}
Applying this to any element of the subspace ${}^0{\cal F}^p$ (which
carries $\rho_p$) we find
\begin{equation}\label{casimir}
  \rho_p(\Delta) = 2p(N+1-p) \;.
\end{equation}
On employing the decomposition (\ref{rhodecompo}), the heat-kernel
averaged correlation function for $\gamma = {\rm e}^{{\rm i} x/N}$
takes the form
\begin{equation}\label{rhodecompoeps}
  \Bl\Omega_U(\gamma)\Br_\eps \defi \e^{-\eps\Delta_U}\Omega_U(\gamma)
  = \sum_{p=0}^{N/2}\e^{-2\eps \, p(N+1-p)}\ \Tr\rho_p(U) \ \frac{\sin
    \big[{x \over 2}(1-\frac{2p-1}{N})\big]} {\sin({x \over 2N})} \;.
\end{equation}
The effect of the averaging procedure is to damp the large--$p$ traces,
which are difficult to estimate from the Gutzwiller trace formula.  In
the above equation the $\eps\to\infty$ behaviour is obvious: all
traces except the trivial one $\Tr\rho_0(U) = 1$ are killed by the
exponential, no matter what the matrix $U$ is.  It is actually not
necessary to set $\eps$ to $\infty$ to get the ${\rm CUE}$
correlation.  Since the irreps $\rho_p$ are unitary, their traces are
bounded by
\begin{displaymath}
  |\Tr\rho_p(U)|\leq {\rm dim}\rho_p = \Tr\rho_p(\mathbb{I}).
\end{displaymath}
The dimensions of the $\rho_p$'s are given in Eq.~(\ref{dimrhop}); for
finite $p$, they are bounded by ${\rm dim}\rho_p\leq~N^{2p}$.  In the
limit $N, p \to \infty$ with $y = p/N$ fixed, Stirling's formula
yields 
\begin{displaymath}
  {\rm dim}(\rho_{p = Ny}) \sim (\pi N)^{-1} \frac{f^\prime(y)}
  {y(1-y)} \, {\rm e}^{2N f(y)} \;,
\end{displaymath}
where the function $f(y) = - y \log y - (1-y) \log (1-y)$ increases
monotonically from $f(0) = 0$ to $f(1/2) = \log 2$.

For any sequence $\{U_N\}_{N \in {\mathbb N}}$, if we tune $\eps$
(possibly varying with $N$) such that
\begin{displaymath}
  \vareps \defi N\eps \gg 1 \;,
\end{displaymath}
all the terms making a significant contribution to
(\ref{rhodecompoeps}) satisfy $p \ll N$.  The $x$--dependence of all
these terms is the same (being given the ${\rm CUE}$ correlation
$x^{-1} \sin(x/2)$), so the averaged correlation will also have this
dependence.  Only the prefactor will depend on the matrices $U_N$
explicitly.  If $\eps$ is increased further to $\vareps\gg\log N$, the
prefactor itself becomes universal.

These statements hold even in the most general case, when the sequence
$\{ U_N \}$ is completely arbitrary.  Therefore, to be able to
differentiate between integrable and chaotic quantum maps, one must
tune the ``disorder strength'' $\vareps$ to smaller values, so that
contributions from the ``high'' traces ${\rm Tr} \rho_p (U_N)$ start
entering into the answer.  To recover the Poisson behaviour for
integrable maps, one actually needs contributions to
(\ref{rhodecompoeps}) coming from the whole region $p \lesssim N/2$.

This puts us in a no-win situation.  On the one hand, we should tune
$\vareps$ to small enough values so that the high traces $p \sim N y$
$(y > 0)$ survive and Poisson behaviour stands a chance to emerge.  On
the other hand, for a chaotic map we have no control over these high
traces (we don't for an integrable map either).

%

For our purposes, the present averaging scheme is probably ``too
algebraic'', as opposed to the semiclassical average presented in
Section \ref{semiclas}.  To motivate this statement in the spirit of
Section \ref{semiclas}, let us compare the spectra of the operators
$\Bl \Ad U \Br$ for the two schemes:
\begin{itemize}
\item The spectrum of $\Bl \Ad U \Br_{\rm semiclas}$
  \emph{qualitatively} depends on the nature of the classical dynamics
  (see Section \ref{semiclas}).  It has a finite gap for a chaotic
  map, whereas eigenvalues accumulate near the unit circle for an
  integrable one.
\item In the isotropic scheme, $\Bl\Ad U \Br_\eps$ is decomposed into
  the irreps $\rho_0(U)\oplus \rho_1(U)$.  Therefore, apart from the
  single eigenvalue unity, $\Bl\Ad U\Br_\vareps$ has the eigenvalues
  $\{\e^{-2\vareps}\e^{\mi(\theta_i-\theta_j)}\}$, where
  $\{\e^{\mi\theta_j}\}$ are the eigenvalues of $U$; the eigenvalue
  $\e^{-2\vareps}$ is $(N-1)$--fold degenerate.  This spectrum is
  qualitatively the same for chaotic versus integrable systems.
\end{itemize}

\subsection{Crossover Poisson-CUE}\label{cross}

We now present an application of the above scheme in the area of
random matrices.  More precisely, we use the isotropic averaging to
build a crossover between the Poisson and ${\rm CUE}$ ensembles, and
we derive the transitional determinant correlation function that
interpolates between the formulas (\ref{omegaresults}).  This
crossover, as well as the method used to compute $\Bl \Omega_U \Br$,
can be compared to the ${\rm GOE} \to {\rm GUE}$ crossover studied in
\cite{smilansky}.

Our crossover is defined as follows.  We start from the Poisson
ensemble, then convolute it with the isotropic (heat kernel) measure
of width $\eps$:
\begin{displaymath}
  \Bl\Omega_U(\gamma) \Br_{{\rm Poisson},\eps} \defi \int_{{\rm U}(N)}
  {\rm d}{\cal P}_{{\rm Poisson}}(U) \int_{{\rm U}(N)} dV K_\eps(V,U)
  \Omega_V (\gamma) \;.
\end{displaymath}
For $\eps = 0$, this is the Poisson ensemble.  In the large--$\eps$
limit, the second integral converges $U$--uniformly to the ${\rm CUE}$
correlation function, so the output $\Bl \Omega_U(\gamma) \Br_{{\rm
    Poisson}, \eps}$ does too.

To calculate the correlation function along the crossover, we will use
the decomposition (\ref{rhodecompoeps}) as in the previous section:
averaging being a linear operation, we only need to replace the
characters $\Tr\rho_p(U)$ by their Poisson averages (see
Eqs.~(\ref{akaverages}) and (\ref{akrhop})):
\begin{displaymath}
  \Bl \Tr\rho_p(U) \Br_{\rm Poisson} = \binom{N}{p}-\binom{N}{p-1}\;.
\end{displaymath}
The asymptotics of these traces in the regime $p, N \to \infty$ with
$y = p/N$ fixed, again follows easily from Stirling's formula:
\begin{equation}\label{poisson}
  \Bl \Tr\rho_p(U) \Br_{\rm Poisson} \sim (2\pi N)^{-1/2}
  \frac{f^\prime(y)} {\sqrt{y(1-y)}} \; {\rm e}^{N f(y)} \;.
\end{equation}
The sum over characters therefore approaches the following integral 
(as $N\to\infty$):
\begin{equation}
  \Bl\Omega_U(\e^{\mi x/N}) \Br_{{\rm Poisson},\eps} \sim \frac{2N^2}
  {\sqrt{2\pi N}} \int_0^{1/2} dy \; \frac{f^\prime(y)} {\sqrt{y(1-
      y)}} \frac{\sin({1 \over 2}x - yx)}{x} \, {\rm e}^{N (f(y) -
    2\vareps y(1-y))} \;.
\end{equation}
In the limit $N \to \infty$, this integral is determined by the saddle
points (rather, the maximum) of $f_\vareps(y) \defi f(y) - 2\vareps
y(1-y)$ on $[0,1/2]$.  Three cases have to be distinguished:
\begin{itemize}
\item If $\vareps < 1$, the boundary point $y = 1/2$ is a maximum of
  $f_\vareps$ and is the only critical point on $[0,1/2]$.  Because of
  the vanishing of the integrand at $y = 1/2$, the saddle--point
  analysis requires some care.  On scales of order ${\cal O}(N^0)$ 
  the result turns out to be independent of $x$:
  \begin{displaymath}
    \Bl \Omega_U(\e^{\mi x/N}) \Br_{{\rm Poisson},\vareps} \sim 2^N
    {\rm e}^{-N\vareps/2} (1-\vareps)^{-3/2} \;,
  \end{displaymath}
  which shows that the Poisson result $2^N$ is retrieved in the limit
  $\vareps \to 0$.  The correlation functions starts depending on $x$
  on scales of order $x \sim {\cal O}(N^{1/2})$.
\item If $\vareps > 1$, the maximum of $f_\vareps$ is situated at the
  point $y_\vareps \in (0,1/2)$ which solves the transcendental
  equation $f_\vareps^\prime (y) = 0$.  The correlation function
  depends on $x \sim {\cal O}(N^0)$ as
  \begin{equation}\label{crossoverexact}
    \Bl \Omega_U(\e^{\mi x/N}) \Br_{{\rm Poisson},\vareps} \propto
    \frac{\sin[x({1 \over 2} - y_\vareps)]}{x} \;.
  \end{equation}
  The flat correlation function has been replaced by an oscillatory
  function, with the period of oscillation being controlled by the
  ``frequency shift'' $y_\vareps$.  When $\vareps$ becomes large, the
  shift vanishes as $y_\vareps\sim \e^{-2\vareps}$, so the ${\rm CUE}$
  correlation function is retrieved.
\item If $\vareps = 1$, the correlation function is ``critical'' (in
  the sense of a phase transition), as the two points $y = 1/2$ and
  $y_\vareps$ coalesce for $\vareps \to 1$ to form a degenerate
  critical point.  In this case the correlation function varies on
  scales $x \sim {\cal O}(N^{1/4})$.
\end{itemize}

\section{Conclusions}

In this paper we have adapted the NL$\sigma$M approach introduced in
\cite{agam,khmel} to the framework of quantized maps on a Hilbert
space of dimension $N\sim\hbar^{-1}$.  We focused on the spectral
determinant correlation function $\Omega_U(\gamma)$ instead of the
pair correlation function, thereby obviating the need to introduce
supersymmetry; we obtained an \emph{exact} expression for the
correlation function as an ordinary integral over a $N^2$--dimensional
complex manifold.  Because the manifold is compact and the integrand
uniformly bounded, no regularization needs to be introduced (unlike in
\cite{agam}).

To estimate this integral we expand the integrand around its saddle
points, first restricting ourselves to the leading--order perturbative
expansion around each point.  Owing to the absence of a large
parameter in front of the effective action, this approximation is
uncontrolled, and the connection between its output and the exact
value of the integral seems fortuitous at best.

Yet, for any matrix $U\in {\rm U}(N)$, we find that the result from
lowest--order saddle--point expansion of the effective action
$S(\gamma,U,Q)$ coincides with the exact correlation function, up to a
global prefactor:
\begin{equation}
  \Omega_U(\gamma)_{\big|{\rm l.o.s.p.~exp.}} = C_N \ 
  \Omega_U(\gamma)_{\rm exact} \;.
\end{equation}
This remarkable coincidence is linked to a cancellation property of
the higher--order terms of the perturbation expansion, which modify
only the prefactor, and is explained by the group-theoretic structure
of the integrand and the Weyl character formula.  Unfortunately, the
expansion is of no use for estimating the correlation function of
quantized maps in the semiclassical limit.

We argue that a decent semiclassical estimate of the correlation
function $\Omega_U(\gamma)$ can only be reached if one takes an
average over a set of unitary matrices in the vicinity of $U$.  To
estimate this averaged correlation, we first average the integrand
$\e^{-S(\gamma,U,Q)}$ over $U$, and then perform the saddle--point
expansion of the output.  Because averaging and saddle--point expansion
are operations that do not commute, this procedure yields an expansion
different from that of the ``individual'' action.  At the same time,
averaging a priori breaks the group-theoretic structure, and with it
the exactness (modulo prefactor) of the leading--order saddle--point
expansion.  Moreover, the explicit computation of saddle points and
their contributions is in general a non--trivial task for a general
averaging scheme.

We have been able to locate the complete set of critical points only
for a certain type of average, namely averaging over all bases of
Hilbert space.  This produces a ${\rm U}(N)$--invariant effective
action, the critical points of which are grouped into submanifolds,
and are independent of the matrix $U$ we started from (as long as its
spectrum is non--degenerate).  Two of these submanifolds are isolated
points; we conjectured that the contributions from these two
``standard'' saddle points, which can be computed explicitly, always
dominate the leading--order saddle--point expansion.

The contributions from these two points are unfortunately ``too
simple'' to constitute a good approximation of the correlation
function, except in some exceptional cases, which we do not truly
understand.  If we average over $U \in {\rm U}(N)$ with the Poisson
measure, the saddle--point result strongly differs from the exact one.
We are thus led to conclude that the leading--order saddle--point
expansion of rotation--averaged effective actions does not yield a
good estimate of the full integral.

What happens in the case of a \emph{local} average, i.e.~when the
weight of the probability measure is concentrated near the quantized
map $U_N$, is unclear.  For one thing, we are only able to exhibit the
two standard saddle points of the averaged action, but there surely
exist many more. 

In the case of the ``semiclassical'' averaging scheme, expansion
around these saddle points yields results similar to those obtained in
\cite{agam}, except that the ``resonances'' we identify are
eigenvalues of a quantum operator.  Yet, these resonances for large
$N$ seem related to the (classical) Ruelle-Pollicott resonances 
\cite{haake01,fishman},
in particular they indicate whether the classical dynamics is chaotic or
integrable.

To connect these resonances with the determinant correlation function
on a rigorous footing, we need two non--trivial assumptions to be
fulfilled.  First, we must assume that the leading--order
saddle--point expansion of the (local average) $S_{\rm semiclas}
(\gamma,U_N,Q)$ makes sense, i.e.~gives a good approximation of the
exact result; the two--loop calculation around $\pm\Sigma_3$ in Section
\ref{loops} seems to support this assumption.  Second, hindered by our
inability to compute the contributions from further saddle points, we
are forced to assume that the full expansion can be truncated to the
two standard saddle points, or at least that this truncation provides
a reasonable approximation.  We presently see no way to prove these
assumptions.

\section*{Appendices}

\subsection*{A. Proof of criticality of the submanifolds ${\cal M}_S$}
\label{criticality}

To prove that the $V$--averaged integrand ${\rm e}^{-S_{V{\rm av}}
  (\gamma,U,Q)}$ is stationary on the submanifolds ${\cal M}_S \subset
{\cal M}_N$, we employ the coherent--state formulation of the
$Q$--integral.  The point $Q_S$ corresponds to the state $| S \rangle
= R({\bf g_\sigma}) | 0 \rangle$, and the points in a neighbourhood of
$Q_S$ may be parametrized as $R({\bf g_\sigma}) | \zeta \rangle$,
where $\zeta$ runs through the (small) $N\times N$ matrices and
$|\zeta \rangle$ is the corresponding coherent state.  The permutation
$\sigma \in \mathfrak{S}_{2N}$ is chosen in such a way as to
interchange the sets $\bar S_1$ and $\tilde S_2 = S_2 + N$, and to
keep $S_1$ and ${\tilde{\bar S}}_2 = \bar S_2 + N$ fixed.

We write the $2N\times 2N$ matrix ${\bf g}_\sigma^{-1} \Gamma {\bf U}
{\bf g}_\sigma$ in the block form $\begin{pmatrix} A &B \\ C &D
\end{pmatrix}$, and first compute the value of the integrand in the 
vicinity of $Q_S$ \emph{before averaging}:
\begin{displaymath}
  \frac{\langle\zeta|R({\bf g}_\sigma^{-1}{\bf \Gamma} {\bf U} {\bf
      g}_\sigma) | \zeta \rangle} {\langle\zeta|\zeta\rangle} = {\rm
    Det}(D) \left( 1 + \Tr(D^{-1}C\zeta + BD^{-1}\zeta^\dagger) +
    {\cal O}(|\zeta|^2) \right) \;.
\end{displaymath}
Then we perform the $V$--average on $U$ (recall that $V \in {\rm
  U}(N)$ acts on $U \in {\rm U}(N)$ by conjugation: $U \mapsto V U
V^{-1}$), and study its output on the right-hand side of the above
equation.  To first order in $\zeta$ and $\zeta^\dagger$, we need the
averages $\Bl {\rm Det}(D) \Br_V$, $\Bl D^{-1}C \ {\rm Det}(D) \Br_V$
and $\Bl BD^{-1} {\rm Det}(D) \Br_V$.  By decomposing the sets $\{1,
\ldots, N \} = S_1 \cup \bar S_1$ and $\{ N+1, \ldots, 2N \} = \tilde
S_2 \cup {\tilde{\bar S}}_2$, the $N \times N$ matrices $B, C, D$ may
be written in block form:
\begin{displaymath}
  B = \begin{pmatrix}\gamma U_{S_1 \bar S_1}&0 \\ 0 &U_{S_2 \bar S_2}
  \end{pmatrix}, \quad C = \begin{pmatrix}\gamma U_{\bar S_1 S_1} 
    &0 \\ 0&U_{\bar S_2 S_2}\end{pmatrix},\quad D =
  \begin{pmatrix} \gamma U_{\bar S_1 \bar S_1}&0 \\ 
    0&U_{\bar S_2 \bar S_2}\end{pmatrix},
\end{displaymath}
where each entry $U_{ss^\prime}$ is a matrix of size $\sharp s \times
\sharp s^\prime$, whose indices take values in the sets $s,s^\prime$.
Thus the $V$--averaged coefficients of the term linear in $\zeta$ 
are the following matrix elements:
\begin{displaymath}
  \Bl {\rm Det} (U_{\bar S_1 \bar S_1}) {\rm Det} (U_{\bar S_2
    \bar S_2}) \left( U_{\bar S_1 \bar S_1}^{-1} U_{\bar S_1
      S_1}^{ \vphantom{-1}} \right)_{ik} \Br_V, \quad \Bl {\rm Det}
  (U_{\bar S_1 \bar S_1}) {\rm Det} (U_{\bar S_2 \bar S_2}) \left(
    U_{\bar S_2 \bar S_2}^{-1} U_{\bar S_2 S_2}^{\vphantom{-1}}
  \right)_{lm} \Br_V \;,
\end{displaymath}
where we have displayed only the dependence on $U$ (and omitted the
$\gamma$--dependence).  We now use the invariance of the Haar measure
$dV$ under (left) multiplication by any unitary matrix and any
diagonal unitary matrix $\delta = {\rm diag} (\delta_1, \dotsc,
\delta_N)$ in particular.  Under such a left translation, the above
matrix elements acquire extra factors $\delta_i/\delta_k$
(resp.~$\delta_l/\delta_m$).  Hence
\begin{displaymath}
  \Bl {\rm Det}(D) (D^{-1}C)_{ik} \Br_V = \Bl {\rm Det}(D)
  (D^{-1}C)_{ik} \Br_V \delta_i/\delta_k \quad\mbox{for any }
  \delta_i, \delta_k \;.
\end{displaymath}
Since $i\in\bar S_1$ and $k\in S_1$ (resp.~$l \in \bar S_2$ and
$m\in S_2$) are never equal and the ratio $\delta_i / \delta_k$ may 
take any value in ${\rm U}(1)$, we conclude
\begin{displaymath}
  \Bl {\rm Det}(D) (D^{-1}C)_{ik} \Br_V = 0 \;.
\end{displaymath}
By the same reasoning, the terms linear in $\zeta^\dagger$ vanish
after $V$--averaging.

We have thus shown that the point $Q_S$ on ${\cal M}_N$ is a critical
point of the $V$--averaged action $S_{V{\rm av}}$, Eq.~(\ref{Vav}).
By the ${\rm U}(N)$--invariance of $S_{V{\rm av}}$, it follows that
the whole submanifold ${\cal M}_S$ is critical for the $V$--averaged
action, no matter what $U$ is.

\subsection*{B. Volumes of the critical submanifolds}

We treat the general case with $\sharp S_1 = \sharp \bar S_2 = N-p$,
$\sharp \bar S_1 = \sharp S_2 = p$, $\sharp(S_1 \cap S_2) = \sharp(
\bar S_1 \cap \bar S_2) = r$, and to build $Q_S$ we use the same
permutation $\sigma$ as in the previous appendix.

The manifold ${\cal M}_{S}$ is given by the set of states $\{R({\bf V
  g_\sigma})|0\rangle ~\big|~ V\in {\rm U}(N)\}$.  These states may be
written (up to normalization) in the form $R({\bf g_\sigma}) | \zeta_V
\rangle$ where the coherent state $| \zeta_V \rangle$ is determined by
the matrix
\begin{equation}\label{zeta_V}
  \zeta_V = \begin{pmatrix}V_{S_1 \bar S_1}^{\vphantom{-1}} V_{\bar
      S_1 \bar S_1}^{-1}&0\\ 0&V_{S_2 \bar S_2}^{\vphantom{-1}}
    V_{\bar S_2 \bar S_2}^{-1}\end{pmatrix} \defi \begin{pmatrix}
    \zeta^{(1)}&0\\0&\zeta^{(2)}\end{pmatrix}
\end{equation}
according to Eqs.~(\ref{gaussian}) and (\ref{cs}).  The block
structure of this matrix derives from the sets $(S_1, S_2)$
vertically, and $(\bar S_1, \bar S_2)$ horizontally.

When $V$ runs through ${\rm U}(N)$, the upper-left matrix $\zeta^
{(1)}$ takes all possible values in $\mathbb{C}^{(N-p)\times p}$.  The
matrix $\zeta^{(2)}$ is not independent of $\zeta^{(1)}$.  For a fixed
$\zeta^{(1)}$, we need to identify the remaining degrees of freedom in
$\zeta^{(2)}$, which is quite easy to do if $\zeta^{(1)}= 0$, i.e.~if
$V$ has the structure $V = {\rm diag}(V_{S_1 S_1}, V_{\bar S_1 \bar
  S_1})$.  The matrices $V_{S_2 \bar S_2}$ and $V_{\bar S_2 \bar S_2}$
in this case block decompose as
\begin{displaymath}
  V_{S_2 \bar S_2} = \begin{pmatrix} V_{12,1\bar 2}&0\\0&V_{\bar
      12,\bar 1\bar 2} \end{pmatrix}, \qquad V_{\bar S_2\bar S_2} =
  \begin{pmatrix} V_{1\bar 2,1\bar 2} &0 \\ 0& V_{\bar 1\bar 2,\bar
      1\bar 2} \end{pmatrix}
\end{displaymath}
where the index $12$ refers to the set $S_1\cap S_2$, etc.  The
degrees of freedom of the lower-right part of $\zeta_V$ are thus two
matrices, $\zeta^{(11)} \defi V_{12,1\bar 2}^{\vphantom{-1}} V_{1\bar
  2,1\bar 2}^{-1} \in \mathbb{C}^{r\times (N-p-r)}$, and $\zeta^{(\bar
  1\bar 1)} \defi V_{\bar 12,\bar 1\bar 2}^{\vphantom{-1}} V_{\bar 1
  \bar 2,\bar 1\bar 2}^{-1}\in\mathbb{C}^{(p-r)\times r}$.  They are
independent of each other, and take all possible values in their
respective vector spaces.  Since the subgroup ${\rm U}(N-p) \times {\rm
  U}(p)$ of ${\rm U}(N)$ acts transitively on the submanifold
$\zeta^{(1)} = 0$ of ${\cal M}_S$, there exists a natural choice of
invariant measure on that submanifold.  It has the factorized form
\begin{displaymath}
  {\rm Det} (1 + {\zeta^{(11)}}^\dagger \zeta^{(11)})^{N-p}
  \prod_{i,j} d^2 \zeta^{(11)}_{ij} / \pi \times {\rm Det} (1 +
  {\zeta^{(\bar 1\bar 1)}}^\dagger \zeta^{(\bar 1\bar 1)})^p
  \prod_{i,j} d^2\zeta^{(\bar 1\bar 1)}_{ij} / \pi \;.
\end{displaymath}
The matrix $\zeta^{(1)}$ parametrizes a coset space ${\rm U}(N)/{\rm
  U}(N-p) \times {\rm U}(p)$, with the corresponding invariant measure
being ${\rm Det}(1 + {\zeta^{(1)}}^\dagger \zeta^{(1)})^N \prod_{i,j}
d^2 \zeta^{(1)}_{ij} / \pi$.  By group invariance arguments, the
volume element of ${\cal M}_S$ (normalized so that it agrees with the
Riemannian measure inherited from the Riemannian manifold ${\cal
  M}_N$) is the product of the measures for $\zeta^{(1)}$, $\zeta^{
  (11)}$, and $\zeta^{(\bar 1 \bar 1)}$ above.  Using this fact and
the result \cite{hua}
\begin{displaymath}
  I(m,n)\defi\int_{\mathbb{C}^{m\times n}}\prod_{i=1}^m\prod_{j=1}^n
  \frac{d^2Z_{ij}}{\pi} {\rm Det}(1+Z^\dagger Z)^{-n-m} =
  \frac{\Gamma(1) \cdots \Gamma(n) \; \Gamma(1) \cdots \Gamma(m)}
  {\Gamma(1) \cdots \Gamma(n+m)},
\end{displaymath}
we obtain the volume of ${\cal M}_S$:
\begin{displaymath}
  {\rm Vol}{\cal M}_S = I(p,N-p) \, I(r,p-r) \, I(r,N-p-r) \;.
\end{displaymath}
A similar integral yields the normalization factor $C_N$ of the
measure $d\mu_N(Z,Z^\dagger)$ on the full manifold ${\cal M}_N$:
\begin{displaymath}
  \frac{1}{C_N}=\int_{\mathbb{C}^{N\times N}} \prod_{i,j=1}^N
  \frac{d^2Z_{ij}}{\pi} {\rm Det}(1+Z^\dagger Z)^{-2N-1} =
  \frac{\Gamma(2)\cdots\Gamma(N+1)}{\Gamma(N+2)\cdots\Gamma(2N+1)}.
\end{displaymath}

\end{document}